\documentclass[prb,eqsecnum,aps,twocolumn,floats,showpacs]{revtex4}
\usepackage{graphicx}
\usepackage{amsmath}

\renewcommand{\eqref}[1]{Eq.~(\ref{#1})}

\newcommand{\non}{\nonumber}
\newcommand{\be}{\begin{equation}}
\newcommand{\ee}{\end{equation}}
\newcommand{\bea}{\begin{eqnarray}}
\newcommand{\eea}{\end{eqnarray}}
\newcommand{\bse}{\begin{subequations}}
\newcommand{\ese}{\end{subequations}}

\newcommand{\lb}{\left[}
\newcommand{\rb}{\right]}
\newcommand{\lp}{\left(}
\newcommand{\rp}{\right)}
\newcommand{\lf}{\left\{}
\newcommand{\rf}{\right\}}
\renewcommand{\k}{\mathbf{k}}
\renewcommand{\r}{\mathbf{r}}

\newcommand{\q}{\mathbf{q}}
\newcommand{\p}{\mathbf{p}}
\newcommand{\bsigma}{\mbox{\boldmath $\sigma$}}
\newcommand{\bpartial}{\mbox{\boldmath $\partial$}}
\newcommand{\HH}{{\cal H}}

\newcommand{\F}{{\cal F}}
\newcommand{\LL}{{\cal L}}

\newcommand{\B}{{\mbox{B}}}
\newcommand{\bF}{\widehat{F}}
\newcommand{\bG}{\widehat{G}}
\newcommand{\tF}{\widetilde{F}}
\newcommand{\tG}{\widetilde{G}}

\renewcommand{\Re}{\mbox{\,Re\,}}
\renewcommand{\Im}{\mbox{\,Im\,}}
\newcommand{\sign}{\mbox{\,sign\,}}
\newcommand{\ts}[1]{\textstyle{#1}}
\newcommand{\ds}[1]{\displaystyle{#1}}
\newcommand{\half}{\textstyle{\frac12}}
\renewcommand{\j}{j}
\newcommand{\pdag}{{\phantom{\dag}}}

\begin{document}
\title{Elastic scattering theory and transport in graphene}
\author{D.~S.~Novikov}
\affiliation{W. I. Fine Institute of Theoretical Physics, 
University of Minnesota, Minneapolis, MN 55455, USA}
\affiliation{Department of Electrical Engineering and Department of Physics, 
Princeton University, Princeton, NJ 08544, USA}
\date{December 28, 2007}


\begin{abstract}

Electron properties of graphene are described in terms of Dirac fermions. 
Here we thoroughly outline the elastic scattering theory 
for the two-dimensional massive Dirac fermions in the presence of an axially 
symmetric potential.
While the massless limit is relevant for pristine
graphene, keeping finite mass allows for generalizations onto situations 
with broken symmetry between the two sublattices, and provides a link to the
scattering theory of electrons in a parabolic band. 
We demonstrate that the Dirac theory requires short-distance regularization
for potentials which are more singular than $1/r$.
The formalism is then applied to scattering off a smooth short-ranged 
potential. Next we consider the Coulomb potential scattering, where  
the Dirac theory is consistent for a point scatterer only for
the effective impurity strength below 1/2. 
From the scattering phase shifts we obtain the exact Coulomb transport cross-section 
in terms of the impurity strength. The results are relevant for 
transport in graphene in the presence of impurities 
that do not induce scattering between the Dirac points in the Brillouin zone.

\end{abstract}
\pacs{81.05.Uw	
      72.10.-d	
      73.63.-b  
      73.40.-c	
}

\maketitle

\section*{\uppercase{Introduction and Outline}}

Graphene, a layer of Carbon atoms arranged in a honeycomb lattice, 
has been long known for its peculiar electronic dispersion,
equivalent to that of massless two-dimensional (2D) Dirac fermions.
This system was first considered in tight-binding approximation by 
Wallace\cite{wallace} and by McClure.\cite{mcclure}
For a long time, graphene monolayer served  
as a low-dimensional toy model where Dirac fermions appear naturally.\cite{semenoff,haldane} Significant interest to this material arose in 1990s 
fueled by the discovery of carbon nanotubes.\cite{Dresselhaus}
The field has experienced an even stronger surge of interest since 2004,
when graphene monolayers were obtained experimentally.
\cite{discovery} The outstanding quality of graphene monolayers and 
few-layered samples is manifest in high mobility resulting in ballistic conductance
on $\mu$m scale, and in quantized Hall effect.
\cite{discovery,novoselov,deheer,zhang,rise}

Recent electron transport measurements \cite{novoselov,deheer,zhang} show 
that the mobility in graphene is approximately independent of the carrier density 
(i.e. conductivity grows proportional to the density). 
The effects of various kinds of the potential disorder on transport in graphene 
have been considered in a number of works.\cite{DVM84,shon-ando,suzuura-ando,peres'06,ando'06,nomura,hwang-adam-dassarma,mccann-ando-altshuler,cheianov-friedel,morpurgo,katsnelson,khveshchenko,ostrovsky,aleiner-efetov,altland,katsnelson-novoselov,hentschel}
Arguably, the density-independent mobility originates mainly due to 
the Coulomb impurities in the substrate.\cite{ando'06,nomura,hwang-adam-dassarma}

The importance of the smooth potential disorder for the transport in graphene 
prompts the development of the scattering theory for the 2D
Dirac fermions, both massless and massive. 
Physically, a nonzero mass can arise due to an external perturbation that 
distinguishes between the sublattices;
recent {\it ab-initio} density functional calculations predict the  
Dirac gap of 53\,meV when placing graphene monolayer on a hexagonal boron nitride 
substrate.\cite{khomyakov} 
Somewhat similar perturbation occurs in bilayer 
graphene,\cite{mcann-falko,nilsson,rotenberg-bilayer} 
although in this example the spectrum is not exactly of the Dirac form.
An interesting possibility for the Dirac gap opening up due to 
the spin-orbit coupling was considered by Kane and Mele.\cite{KM'05}

The purpose of this work is to thoroughly outline the elastic scattering
theory for 2D Dirac fermions in the axially-symmetric potentials.
Such a 2D formalism is built essentially
following Ref.~\onlinecite{Landau4}, the classic reference for scattering in
3D Dirac systems. 
The connection to the transport in graphene in the presence
of potential centers whose field is smooth on the lattice scale is established
via the transport cross section.

We start from the basic facts about the 2D Dirac model, and 
write the normalized spinor plane and spherical 
waves (Sec.~\ref{sec:free}). In Sec.~\ref{sec:scattering} 
we study the properties of the radial solutions for the 2D Dirac spinors, define 
the scattering phase shifts, and link them to the differential
and transport cross sections.
We also derive the Born approximation for the 2D Dirac spinors, as well as outline
analytical properties of the radial solutions on the complex energy plane.

One of the observations made in Sec.~\ref{sec:scattering} is that the 
Dirac problem, both massive and massless, requires a short-distance regularization
whenever the external potential is more singular than $1/r$.
Classically, this corresponds to falling into the potential center. 
For such singular potentials the purely Dirac formalism is inapplicable,
and the lattice scale physics starts playing a role.

In Sec.~\ref{sec:slow} we consider scattering off a potential
localized within a finite radius that exceeds the lattice constant
but is smaller than the particle wavelength.

In Sec.~\ref{sec:coul} we focus on one of the most important scattering problems
for graphene --- that for the potential $U=-\hbar v \times \alpha/r$.  
The latter problem has so far been 
treated in the Born approximation.\cite{ando'06,nomura,hwang-adam-dassarma}
The exact solution is presented in detail for both massless and massive cases, 
and for both signs of the impurity charge $\alpha$. The asymptotic behavior of 
the scattering solutions and scattering phase shifts are studied, with the 
particular attention paid to the ``ultrarelativistic'' 
and ``nonrelativistic'' limits,
relevant correspondingly for the pristine graphene, and for the graphene layer with 
broken sublattice symmetry, or for the electrons in a semiconductor with 
a parabolic band.

The $1/r$ problem deserves a special consideration, as it is a borderline
case for the falling into the potential center. 
It has been known\cite{khalilov} that the solutions in the $1/r$ potential
become singular whenever the ``fine structure constant'' $\alpha>\alpha_c=1/2$,
i.e. at even smaller value than that in 3D ($\alpha_c=1$).
In this respect the physics of the Coulomb impurity in graphene, depending
on the dielectric environment, may correspond to the ``supercritical''
relativistic heavy atom (that with $Z>137$).
\cite{Pomeranchuk,Gershtein,Popov,ZP}

Finally, In Sec.~\ref{sec:transport} we use the exact scattering phases
to calculate the transport cross section for the subcritical Coulomb impurity,
and compare the exact result with the Born approximation.
Our main finding there is that, for a given carrier charge, the attracting 
impurity scatters more effectively than the repelling one.
This should be contrasted with known exact 2D and 3D nonrelativistic scattering
results in a $1/r$ potential, where such an asymmetry does not take place.

\section{Free electrons in graphene}
\label{sec:free}

\subsection{Model}

The key features of electron dispersion in an ideal graphene monolayer
can be summarized as follows.\cite{Dresselhaus} 
With the two cites per unit cell, graphene's $\pi$-electron band has 
the two inequivalent points in the Brillouin zone, 
at which the electron and hole subbands just barely touch.
At these so-called Dirac points $K$ and $K'$, the carrier dispersion is linear
and electron-hole symmetric, $\epsilon(\p) \propto \pm p$. 
Separately at the $K$ and $K'$ points, 
the wave function in the effective-mass approximation
has a spinor structure, its two components corresponding
to the two sublattices. 
This spinor obeys the massless Dirac equation.
The low-energy states near the $K$ and $K'$ points are decoupled in a pristine
graphene monolayer. Provided the material is subject to external fields
that are adiabatic on the lattice scale, the low-energy properties 
can be understood in terms of the $N_f=2_{\rm spin}\times 2_{\rm valley}=4$
independent Dirac fermion polarizations.

Near a given Dirac point, the free effective-mass Hamiltonian 
\be \label{H0}
\HH_0 = -i\hbar v (\tau_3\sigma_1 \partial_x + \sigma_2 \partial_y) 
+ \Delta \sigma_3 \,,
\ee
where $v\approx 1\times 10^6\,$m/s is the graphene Fermi velocity, 
$\sigma_{1,2,3}$ are the Pauli matrices that act in the spinor space corresponding
to the two sublattices of a honeycomb lattice,
while $\tau_3=\pm 1$ distinguishes between the $K$ and $K'$ Dirac points.
Everywhere in this work we consider the
dynamics on the scale much larger than the graphene's lattice constant
and neglect scattering between the Dirac points. Hence,
without loss of generality, we set $\tau_3=+1$ in what follows.

In the Hamiltonian (\ref{H0}) we also introduced  
the gap (the Dirac mass)
\be \label{M}
\Delta \equiv Mv^2 \,.
\ee
Although this term is absent by symmetry in an ideal graphene monolayer, 
it can become imporant when the symmetry is reduced.
Without loss of generality here we set $M>0$
(working at zero magnetic field, 
we are not concerned with the parity anomaly effects\cite{semenoff,haldane}).


The Hamiltonian (\ref{H0}) corresponds to the Lagrangian [$\hbar=v=1$, $\tau_3=+1$]
\be \label{L0}
\LL_0 = \bar\psi \lp i\gamma^\mu \partial_\mu  - M \rp \psi \,,
\quad \bar\psi \equiv \psi^\dag \gamma^0 \,,
\ee
where the Dirac matrices
$\gamma^0=\sigma_3$, $\gamma^1=i\sigma_2$ and $\gamma^2=-i\sigma_1$,
such that $\{ \gamma^\mu, \gamma^\nu \} = 2g^{\mu\nu}$ with 
$g^{\mu\nu}={\rm diag\,}(1, -1, -1)$. 
In this ``relativistic'' notation the Lorentz-invariant fermion current is 
\be
J^\mu=\bar\psi \gamma^\mu\psi = (\rho, \ {\bf J}) \,, 
\ee
where the number density and the current are 
\be \label{J-rho}
\rho = \psi^\dag \psi \,, \quad 
{\bf J} = \psi^\dag \bsigma \psi \,,
\ee
with
$\bsigma=(\sigma_1, \ \sigma_2)$.

\subsection{Spinor plane waves}

Consider the eigenproblem $\HH_0\psi = \epsilon \psi$, Eq.~(\ref{H0}), 
where we represent the two-component spinor
\be \label{psi-u}
\psi(\r) = \begin{pmatrix} \varphi \\ \chi \end{pmatrix} .
\ee 
The components of $\psi$ satisfy [we set $\hbar = v = 1$ in what follows] 
\be \label{matrix-phi-chi}
\begin{pmatrix} M & p_x - ip_y \\ p_x + ip_y & -M \end{pmatrix} 
\begin{pmatrix} \varphi \\ \chi \end{pmatrix} 
= \epsilon \begin{pmatrix} \varphi \\ \chi \end{pmatrix} ,
\ee
where $p_x=-i\partial_x$ and $p_y=-i\partial_y$. In the plane wave basis
$(\varphi \ \chi)^T = u_{\epsilon,\p} e^{i\p\r}$
the differential operators become components of the momentum eigenvalue $\p$,
yielding the relativistic dispersion
\be \label{ep}
\epsilon 
= \pm \sqrt{p^2 + M^2} \,, \quad p = \sqrt{p_x^2 + p_y^2} \,,
\ee
where $\pm$ distinguishes between the particle and hole sectors.
The conventional normalization ``one particle in a unit volume'' \cite{Landau4}
$J^{\mu}=p^\mu/\epsilon=(1, {\bf v}_\p)$, where 
${\bf v}_\p=\partial \epsilon/\partial\p = \p/\epsilon$ is the velocity, 
requires $\bar \psi_{\epsilon,\p} \psi_{\epsilon,\p} \equiv M/\epsilon$ 
or, equivalently, $\psi_{\epsilon,\p}^\dag \psi^\pdag_{\epsilon,\p}=1$, 
yielding
\be \label{u-plane}
\psi_{\epsilon; \,\p} = u_{\epsilon; \,\p} e^{i\p\r}\,, \quad
u_{\pm |\epsilon|; \,\p} = {w\over \sqrt{2|\epsilon|}}
\begin{pmatrix}
\sqrt{|\epsilon+M|}
\\ 
\pm \sqrt{|\epsilon-M|} e^{i\theta_\p}
\end{pmatrix} .
\ee
Here $\theta_\p = {\rm arg\,} (p_x + ip_y)$, and the upper and lower signs 
refer to the electron  ($\epsilon>M$) and hole ($\epsilon<-M$) 
parts of the spectrum, $\pm \equiv \sign\epsilon$.  
The absolute values under the square roots are introduced to describe both
sectors. The factor $w=e^{i\phi}$ is an overall phase 
that has a meaning of the ``nonrelativistic''
particle's wave function in the rest frame, $\epsilon = M$.

\subsection{Spinor spherical waves}

For the purpose of developing the scattering theory, below we introduce 
the spherical wave basis of eigenstates of the problem (\ref{H0}).

First recall that the 2D nonrelativistic scalar particle with fixed absolute 
value of the momentum $p=|\p|$ and fixed projection 
$m=0, \pm 1, \pm2, ...$ of angular momentum on  the $z$-axis (perpendicular to 
the plane) is described by the spherical wave
\be
\Psi_{pm}(\r) = \Phi_m(\theta) R_{pm}(r) \,.
\ee
Here the angular harmonics 
\be \label{Phi-m}
\Phi_m = {1\over \sqrt{2\pi}} \, e^{im\theta} \,, 
\quad m=0,\pm 1, \pm 2, ...  \,,
\ee
and the radial functions $R_{pm}(r)$ 
satisfy the radial Schr\"odinger equation
\be \label{R-free-schr}
-{1\over r}  {d\over dr} \lp r {d\over dr} R_{pm}\rp + {m^2\over r^2} R_{pm} 
= p^2 R_{pm} \,,
\ee
which reduces to the Bessel equation
\be \label{bessel}
\rho^2 R''_{\rho\rho} + \rho R'_\rho + (\rho^2 - m^2)R = 0 \,, \quad \rho=pr \,.
\ee
The solutions are the Bessel functions $J_m(\rho)$ and the Neumann functions 
$Y_m(\rho)$, whose asymptotic behavior 
\begin{subequations}
\label{asympt-large-JY}
\bea 
J_m(\rho\gg m) &\simeq& 
\sqrt{2\over \pi \rho} \cos\lp \rho - {m\pi\over 2} - {\pi\over 4}\rp , \quad
\\
Y_m(\rho\gg m) &\simeq& 
\sqrt{2\over \pi \rho} \sin\lp \rho - {m\pi\over 2} - {\pi\over 4}\rp . \quad
\eea
\end{subequations}
Their corresponding short-distance behavior is
\bse \label{bessel-short}
\be 
J_m(pr) \sim {1\over m!} \lp \frac{pr}2 \rp^m 
\ee
and 
\be
Y_m(pr) \sim \left\{
\begin{matrix}
-{\Gamma(m)\over \pi} \lp {2\over pr}\rp^m , \ \  m > 0 \,; 
\\
{2\over \pi} \ln\lp\gamma_E pr/ 2\rp , \ \ m=0 \,,
\end{matrix}
\right.
\ee
\ese
where $\ln \gamma_E\simeq 0.577...$ is the Euler's constant.
[For $m<0$, use $J_{-m}=(-)^m J_m$ and $Y_{-m}=(-)^m Y_m $.]
Thus for the wave regular at $r=0$, one chooses $R_m\propto J_m(pr)$, normalized
according to $\int_0^\infty \! rdr \, R_{pm}R_{p'm'} = 2\pi\delta_{mm'}\delta(p-p')$:
\be \label{asympt-large}
R_{pm} = \sqrt{2\pi p}\times  J_m(pr) \sim {2\over \sqrt{r}} 
\cos\lp pr - {m\pi\over 2} - {\pi\over 4}\rp .
\ee

Coming back to the relativistic case, one notes that both the isospin 
$\half{\bsigma}$ and the angular momentum 
$\hat l_z = -i \partial_\theta = -i(x\partial_y - y\partial_x)$ do not commute with
the Hamiltonian (\ref{H0}): 
\[
[\hat l_z, \ \HH_0] = i\bsigma\times\p \,, \quad
[ \half\hat\sigma_z, \ \HH_0] = -i\bsigma\times\p \,,
\]
thus a state cannot be characterized by their values. 
(In fact, the spherical spinor components will have different eigenvalues of 
$\hat l_z$.)
The conserved quantity is the ``isospin-orbital'' momentum around the $\hat z$-axis,
\cite{DVM84} 
\be \label{Kappa}
\hat\j = \hat l_z + \half \sigma_z \,, 
\quad \hat l_z = -i\partial_\theta \,.
\ee
Also, similar to the 3D case, \cite{Landau4}
the {\it parity} of a state is conserved: Under inversion 
$\r\to -\r$ [i.e. $\theta \to \theta + \pi$ for the polar angle, 
and $\psi\to \gamma^0\psi$], the spinor components (\ref{psi-u}) transform as 
$\varphi(\r) \to \varphi(-\r)$ and $\chi(\r) \to -\chi(-\r)$. The spinor 
$\psi_{pm}$ will have the definite parity $(-)^m$ if its components 
\be \label{def-FG}
\psi_{pm}(\r) = 
\begin{pmatrix} F_{pm}(r) \Phi_m(\theta) \\ 
iG_{pm}(r) \Phi_{m+1}(\theta) \end{pmatrix} 
\ee
have the angular parts correspondingly with $l_z=m$ and $l_z=m+1$.
The factor of $i$ here is chosen for later convenience.

Consider now the radial parts $F_{pm}(r)$  and $G_{pm}(r)$
assuming $|\epsilon|>M$. 
From the relations (\ref{matrix-phi-chi}), 
it follows that
\bse \label{R-free-dir}
\bea
-{1\over r}  {d\over dr} \lp r {d F_{pm}\over dr} \rp 
+ {m^2\over r^2} F_{pm} &=& p^2 F_{pm} \,, \quad \quad 
\\
-{1\over r}  {d\over dr} \lp r {d G_{pm}\over dr}\rp 
+ {(m+1)^2\over r^2} G_{pm} &=& p^2 G_{pm} \,. \quad \quad
\eea
\ese
The equations (\ref{R-free-dir}) are of the 
radial Schr\"odinger form, Eq.~(\ref{R-free-schr}).
Thus $F_{pm} = A R_{pm}(r)$ and $G_{pm} = B R_{p,m+1}(r)$.
For the spinor wave regular at the origin, one chooses the radial functions
in the form (\ref{asympt-large}).
To find $A$ and $B$,
we consider the limit $pr\to \infty$, when the wave function is 
approximately a plane wave in the direction of $\hat \r$.
Using the asymptotic behavior (\ref{asympt-large}) and the relations (\ref{u-plane})
between the components of the plane wave, find 
$B/A=\pm \sqrt{|\epsilon - M|/|\epsilon + M|}$,
where $\pm \equiv \sign \epsilon$.
Requiring the overall normalization
$\int\! d^2\r \, \psi^\dag_{pm}\psi^\pdag_{p'm'} = 2\pi \delta_{mm'} \delta(p-p')$,
 obtain the spinor spherical wave 
\be \label{sph}
\psi_{pm}(\r) = {1\over \sqrt{2|\epsilon|}}
\begin{pmatrix}
\sqrt{|\epsilon + M|} R_{pm}(r) \Phi_m(\theta) 
\\
\pm i\sqrt{|\epsilon - M|} R_{p, m+1}(r) \Phi_{m+1}(\theta)
\end{pmatrix} , 
\ee
whose parity is $(-)^m$.
Here $R_{pm}(r)=\sqrt{2\pi p} \,J_m(pr)$ for the spinor regular at $r=0$,
$R_{pm}(r)=\sqrt{2\pi p} \,Y_m(pr)$ for the spinor singular at $r=0$,
and same applies for $R_{p,m+1}(r)$. 
The spinor (\ref{sph}) is also an eigenstate of the operator (\ref{Kappa})
with the eigenvalue 
\be \label{kappa}
\j=m+\half \,.
\ee
Below we will often use the eigenvalue (\ref{kappa}) instead of 
the orbital number $m$ to label states, eigenvalues or phase shifts;
e.g. for the spinors (\ref{sph}), $\psi_{pm}(\r) \equiv \psi_{pj}(\r)$.

\section{Potential scattering}
\label{sec:scattering}

\subsection{Equations for the radial functions}
\label{sec:rad}

We consider elastic 
scattering off an axially-symmetric external scalar potential $U(r)$. 
The Hamiltonian 
\be \label{H0+U}
\HH = -i (\sigma_1 \partial_x + \sigma_2 \partial_y) + M \sigma_3 
+ U(r)\,.
\ee
In the spinor components (\ref{psi-u}), Eq.~(\ref{H0+U}) reads 
\be \label{phi-chi-U}
\begin{matrix}
(\epsilon - M - U)\varphi = (p_x-ip_y)\chi \,, \\ 
(\epsilon + M - U)\chi = (p_x+ip_y) \varphi \,, 
\end{matrix}
\ee
where $(p_x, \ p_y) = (-i\partial_x, \ -i\partial_y)$ are differential operators.

The crucial symmetry of the problem (\ref{H0+U}) 
is the conservation of the total orbital momentum (\ref{Kappa}), since
\be \label{jz-comm}
[\hat l_z, \ \HH] = -[\half\hat\sigma_z, \ \HH]= i\bsigma\times\p  \quad
\Rightarrow \quad  [\hat j_z, \ \HH] = 0 
\ee
for any axially-symmetric $U(r)$. This property allows us to work in 
the spherical basis of the form (\ref{def-FG}).
Taking the spinor (\ref{psi-u}) in the form (\ref{def-FG}),
and using $\Phi_{m+1}= e^{i\theta}\Phi_m$, and
$p_x\pm ip_y=e^{\pm i\theta}(-i\partial_r \pm \frac1r \partial_\theta)$,
where $\theta=\arg(x+iy)$,
obtain the following equations for the radial functions $F$ and $G$:
\begin{subequations}
\label{FG-m}
\bea
\label{F-m}
{dF\over dr} - {m\over r} F + (\epsilon + M - U) G &=& 0 \,, \\ 
\label{G-m}
{dG\over dr} + {m+1\over r} G - (\epsilon - M - U) F &=& 0 \,.
\eea
\end{subequations}
Everywhere here it is implied that the functions $F$ and $G$ 
correspond to the angular momentum (\ref{kappa}), e.g. 
$F\equiv F_m \equiv F_j$; the index $j$ or $m$ will be often suppressed for brevity.
Eqs.~(\ref{FG-m}) have been derived by DiVincenzo and Mele \cite{DVM84}
for the case $M=0$.
In the absence of the potential, $U\equiv 0$, Eqs.~(\ref{FG-m})
are equivalent to Eqs.~(\ref{R-free-dir}).

It is often convenient to represent Eqs.~(\ref{FG-m}) in 
a more symmetric form, using the eigenvalue (\ref{kappa}),
\begin{subequations}
\label{FG-kappa}
\bea
\label{F-kappa}
(F\sqrt{r})'_r - {\j\over r} (F\sqrt{r}) + (\epsilon+M-U)(G\sqrt{r})&=& 0, \quad 
\\
\label{G-kappa}
(G\sqrt{r})'_r + {\j\over r} (G\sqrt{r}) - (\epsilon-M-U)(F\sqrt{r})&=& 0. \quad 
\eea
\end{subequations}

Eqs.~(\ref{FG-m}) and (\ref{FG-kappa}) are valid both for the continuous and for 
the discrete spectrum (present for $M\neq 0$).
In the massless limit $M\to 0$, Eqs.~(\ref{FG-m}) and (\ref{FG-kappa}) 
acquire the following symmetry: For any $U(r)$,
if a pair $(F, G)$ is the solution for a given $\j$, 
then the pair $(G,-F)$ is the corresponding solution 
for $\j \to -\j$, i.e.
\be \label{-j}
M=0: \quad F_{-j} = G_j \,, \quad G_{-j} = -F_j \,.
\ee
This symmetry is also present for the 3D massless Dirac fermions
[Ref.~\onlinecite{Landau4}, Sec.~38].

The asymmetry between the spinor components associated with 
the finite Dirac mass $M$ is stressed by rescaling
the radial functions in accord with Eq.~(\ref{sph}),
\be \label{tilde-FG}
|\epsilon|>M: \quad 
\begin{matrix}
F\sqrt{r} &=& \sqrt{\pm (\epsilon+M)} \bF \,,  \\
G\sqrt{r} &=& \pm \sqrt{\pm (\epsilon-M)} \bG \,, 
\end{matrix}
\ee
where $\pm$ is $\sign \epsilon$. Then Eqs.~(\ref{FG-kappa}) take the form
\begin{subequations}
\label{tilde-FG-kappa}
\bea
\label{tilde-F-kappa}
\bF'_r - {\j\over r} \bF 
+ p\lb 1- {U\over \epsilon + M}\rb \bG&=& 0, \quad \quad
\\
\label{tilde-G-kappa}
\bG'_r + {\j\over r} \bG 
- p\lb 1- {U\over \epsilon - M}\rb \bF&=& 0. \quad \quad
\eea
\end{subequations}
Here $p=\sqrt{\epsilon^2-M^2}$.
The same can be done for the discrete spectrum, 
\be \label{tilde-FG-dis}
|\epsilon|<M: \quad 
\begin{matrix}
F\sqrt{r} = \sqrt{M+\epsilon}\, \bF \,, \\
G\sqrt{r} = \sqrt{M-\epsilon}\, \bG \,. 
\end{matrix}
\ee
Introducing $\lambda \equiv \sqrt{M^2-\epsilon^2}$,
the corresponding equations 
\begin{subequations}
\label{tilde-FG-kappa-dis}
\bea
\label{tilde-F-kappa-dis}
\bF'_r - {\j\over r} \bF 
+ \lambda \lb 1- {U\over M+\epsilon}\rb \bG = 0, \quad \quad
\\
\label{tilde-G-kappa-dis}
\bG'_r + {\j\over r} \bG 
+ \lambda \lb 1+ {U\over M-\epsilon}\rb \bF = 0. \quad \quad
\eea
\end{subequations}

Finally, we reduce the system of first-order equations, 
say Eqs.~(\ref{tilde-FG-kappa}),
to an equivalent second-order equation. 
The latter can be written either 
for $\bF$ or for $\bG$, as follows:
%
%
\be \label{barF}
\bF'' + {U' \lb \bF' - {j\over r}\bF\rb  \over \epsilon+M-U}  
+\lb (\epsilon-U)^2 - M^2 + {j-j^2\over r^2} \rb \bF = 0
\ee
[the corresponding equation for $\bG$ would have $j\to -j$].
Eq.~(\ref{barF}) is reduced to the familiar Schr\"odinger form 
\be \label{schr}
\Psi'' + 2\lb E - V(r)\rb \Psi = 0 \,, \quad E = p^2/ 2\,,
\ee
via the substitution $\bF = \sqrt{\epsilon+M-U} \, \Psi$. 
Similar to the 3D case, the potential $V=V_1+V_2$
splits into the Klein-Gordon part $V_1$, and the part $V_2$ responsible for 
the Dirac ``spin'' effects:
\bea 
V_1 &=&\epsilon U(r) -\frac12 U^2 + {j^2-j\over 2r^2} \,, \\
V_2 &=& \frac14 \lb {U''\over \epsilon+M-U} 
+ \frac32 \lp{U'\over \epsilon+M-U}\rp^2 
+ {2\frac{j}{r} U' \over \epsilon+M-U}\rb .
\non
\eea
It should be clear that the ``spin''-orbit coupling coming from the potential $V_2$
has nothing to do either with the real SU(2) spin of electrons, or 
with having the two Dirac points. Rather, it is a consequence of a two-component
spinor structure of the electron wave function due to the existence of the two
sublattices in a honeycomb lattice. We also note that the Schr\"odinger 
``energy variable'' $E$ in Eq.~(\ref{schr}) has the dimension of [energy]$^2$,
same as that of the potential $V$. In effect, Eq.~(\ref{schr}) is a square
of the original Dirac problem (\ref{H0+U}), hence 
the original potential $U$ alone, and even its sign, do not have a transparent 
meaning in the problem (\ref{schr}).

\subsection{Short-distance behavior: \\ Limitations on the Dirac description}
\label{sec:limit-dirac}

Consider the potential $U(r)$ that at $r\to 0$ is more singular than $1/r$.
In this case, for small $r$, Eqs.~(\ref{F-m}) and (\ref{G-m}) take the form
\be \label{eqs-short}
F'_r -UG = 0 \quad {\rm and\ } \quad G'_r +UF = 0 \,,
\ee
whose solutions are 
\be \label{FG-short}
F = C\sin\ts{\lp\int^r \! U dr + \delta\rp } , \quad 
G = C\cos\lp\int^r \! U dr + \delta\rp 
\ee
with constant $C$ and $\delta$.
These functions strongly oscillate and have no limit for $r\to 0$. In
the nonrelativistic case this situation would be equivalent to falling into 
the source of the potential: namely, such a potential
allows for infinitely deep-lying bound states.\cite{Landau3} 
Physically, a sufficiently singular potential in a (massive) 
relativistic system causes the Dirac vacuum breakdown 
(the Schwinger effect).\cite{schwinger} 
Such a singular attractive potential will
be responsible for the free emission of electron-hole pairs; 
if the potential is attractive, electrons would then bind to it, 
while holes will be pushed to infinity.\cite{Landau4}
If the potential is repulsive, it will push away 
the electrons and bind holes instead. 

In the Schr\"odinger case falling into the potential center 
first occurs for the $1/r^2$ singularity.\cite{Landau3} 
It is not surprising that the Dirac problem is more sensitive to 
singular behavior at short distances, as it can be roughly thought of as a 
``square root'' of the Schr\"odinger equation. 

As a result of these simple considerations, for both repulsion and attraction, 
the potentials that are more singular than $1/r$ for $r\to 0$ 
cannot be correctly considered
within the low-energy effective Dirac theory (\ref{H0+U}). 
In this case the exact eigenstates have to be 
determined on the length scale of the graphene lattice,
where the long-wavelength description (\ref{H0+U}) breaks down. 
Such a situation, where the effect of the lattice cannot be simply incorporated
by means of the effective-mass description, is reminiscent of that for the   
deep-lying impurity levels in the middle of the band gap in a semiconductor,
where the effective-mass theory is inapplicable from the outset. 

\subsection{Scattering amplitude and cross section}

Below we develop the elastic scattering theory for the 2D Dirac fermions
in the presence of the axially-symmetric potential $U(r)$.
Our goal is to express the scattering amplitude 
and the cross section in terms of the scattering phase shifts
for the spinor spherical waves of the form (\ref{sph}).

First we recall that in the nonrelativistic case, 
with the incident flux along the $\hat{\bf x}$-direction,
the 2D wave function has the asymptotic form 
[our notation follows Ref.~\onlinecite{Landau3}]
\be \label{def-f}
\Psi \simeq e^{ipx} + {f(\theta)\over \sqrt{-ir}} \times e^{ipr} \,,
\ee
where $f$ is the 2D scattering amplitude,
and the factor $\sqrt{-i} = e^{-i\pi/4}$ is introduced for further convenience.
The differential and the total cross sections, 
that have the dimensionality of length, are\cite{Landau3}
\be \label{def-sigma}
{d\Lambda\over d\theta} = |f(\theta)|^2 \,, \quad
\Lambda = \oint \! |f(\theta)|^2 \, d\theta \,.
\ee
[We have denoted the scattering cross section by $\Lambda$ since the letter $\sigma$
is commonly reserved for the conductivity.]
One way to find the scattering amplitude $f$ is to represent the wave function $\Psi$ 
in the spherical wave basis,
$\Psi = \sum_m A_m R_{pm}(r) \Phi_m(\theta)$,
and to consider the Schr\"odinger equation 
\be \label{schr-m}
{1\over r}(rR_{pm}')' 
+ \lb p^2 - {m^2\over r^2} - {2M U(r)\over \hbar^2}\rb R_{pm} = 0
\ee
for each of the radial components $R_{pm}$.
The scattering phase shifts $\delta_m$ are then 
defined by the asymptotic form of the solutions of Eq.~(\ref{schr-m}):
\be \label{R-asympt}
R_{pm}(r) \simeq {2\over \sqrt{r}} 
\cos\lp pr - {m\pi\over 2} -{\pi\over 4}+\delta_m \rp .
\ee
Using the decomposition of the plane wave
\be \label{plane-wave}
e^{ipx} = \sum_{m=-\infty}^\infty i^m J_m(pr) e^{im\theta} \,, 
\ee
together with the definition (\ref{def-f}), find 
\be \label{Am}
A_m=i^m p^{-\frac12}  e^{i\delta_m} \,,
\ee
and
\be \label{f}
f(\theta) = \frac1{i\sqrt{2\pi p}} \sum_{m=-\infty}^\infty (S_m-1) e^{im\theta} \,,
\quad S_m \equiv e^{2i\delta_m} \,.
\ee
From Eq.~(\ref{f}), the total cross section (\ref{def-sigma}) follows: 
\cite{Stern-Howard,Barton}
\be \label{sigma-delta}
\Lambda = {4\over p} \sum_{m=-\infty}^\infty \sin^2\delta_m \,,
\ee
and the momentum-relaxation (transport) cross section
\be \label{sigma-tr}
\Lambda_{\rm tr} = \oint\! d\theta\, (1-\cos\theta)|f(\theta)|^2 
= {2\over p} \sum_{m=-\infty}^\infty \sin^2(\delta_{m+1}-\delta_m) \,.
\ee
With our definition of $f$, the 2D optical theorem is then
\be \label{opt-thm}
\Lambda =  \sqrt{8\pi/p} \times {\rm Im\, } f(0) \,.
\ee

Turning to the Dirac case with the Hamiltonian (\ref{H0+U}),
the asymptotic form for the spinor wave function is
\be \label{def-f-spinor}
\psi = u_{\epsilon,p\hat{\bf x}} e^{ipx} 
+ {f(\theta)\over \sqrt{-ir}} \times u_{\epsilon,\p_\theta} e^{ipr}\,,
\ee
where $\p_\theta=p(\cos\theta,\sin\theta)$ defines the direction of scattering,
and $u_{\epsilon,\p}$ is the normalized plane wave amplitude (\ref{u-plane}).
Since, according to Eq.~(\ref{J-rho}), the scattered current 
\be \label{Js}
{\bf J}_{\rm scatt} = {|f|^2\over r} 
u_{\epsilon,\p_\theta}^\dag \bsigma u_{\epsilon,\p_\theta} 
= {|f|^2\over r} {\p_\theta \over \epsilon} \,,
\ee
and the incident current  ${\bf J}_{\rm in} = p\hat{\bf x}/\epsilon$, 
it follows that the scattering 
amplitude $f$ is analogous to that in the nonrelativistic case, with the 
cross section given by Eq.~(\ref{def-sigma}).
Similarly, one defines the scattering phase shifts $\delta_j$ via the asymptotic 
form of the radial wave functions of the spherical spinor (\ref{def-FG}), 
\be \label{def-FG-j}
\psi_j(\r) = 
\begin{pmatrix} F(r) \Phi_{j-1/2}(\theta) \\ 
iG(r) \Phi_{j+1/2}(\theta)\end{pmatrix}
\ee
[here we relabeled $\psi_m \to \psi_j$ using (\ref{kappa})].
The wave functions $F$ and $G$ are determined by 
Eqs.~(\ref{FG-m}), or Eqs.~(\ref{FG-kappa}).
Their asymptotic behavior 
should then be compared to that of the free spherical spinor (\ref{sph})
with $R_{pm}(r)$ regular at $r=0$,
%
%
\begin{subequations}
\label{FG-asympt}
\bea \label{FG-asymptF}
F  & \simeq &  
{2\over \sqrt{r}}\sqrt{|\epsilon+M| \over 2|\epsilon|}
\cos\lp pr - {j\pi\over 2} + \delta_j \rp, \quad\quad\quad
\\
\label{FG-asymptG}
G & \simeq & \pm  {2\over \sqrt{r}} \sqrt{|\epsilon-M| \over 2|\epsilon|}
\sin\lp pr - {j\pi\over 2} + \delta_j \rp. \quad\quad\quad
\eea
\end{subequations}
The spinor wave function (\ref{def-f-spinor}) is represented in the basis
(\ref{def-FG}) as $\psi = \sum_j A_j \psi_{j}(\r)$, where the 
coefficients $A_j$, expressed in terms of the phase shifts $\delta_j$ introduced
in Eq.~(\ref{FG-asympt}), are given by the ``nonrelativistic''  
Eq.~(\ref{Am}), $A_j = i^{j-1/2}p^{-1/2} \exp(i\delta_j)$.
The scattering amplitude, the cross section and the optical theorem 
directly follow, cf. Eqs.~(\ref{f}), (\ref{sigma-delta}) and (\ref{opt-thm}) 
correspondingly:
\bea \label{f-rel}
f(\theta) &=& \frac1{i \sqrt{2\pi p}} \sum_{j=\pm \frac12, \pm \frac32, ...} 
\!\! (S_j-1)e^{i(j-1/2)\theta} \quad\quad\quad \\
\label{S-j}
\mbox{where\ \ } S_j &\equiv& e^{2i\delta_j} \,;
\\ 
\label{sigma-delta-rel}
\Lambda &=& {4\over p} \sum_{j=\pm \frac12, \pm \frac32, ...} \sin^2\delta_j \,,
\\ 
\label{lambda-tr-gen}
\Lambda_{\rm tr} &=& {2\over p} \sum_{j=\pm \frac12, \pm \frac32, ...}
\sin^2\lp \delta_{j+1}-\delta_j \rp .
\eea

Finally, consider the most common massless case, 
where we derive an important property
\be \label{j=-j}
M=0: \quad 
\delta_{-\j}=\delta_\j \,.
\ee
For that we apply the symmetry (\ref{-j}) to the asymptotic form (\ref{FG-asympt})
and use Eqs.~(\ref{asympt-large-JY}) 
[noting that $\delta_j$ are defined modulo $\pi$, since the observable 
quantities are the $S$-matrix elements (\ref{S-j})].
The property (\ref{j=-j}) ensures that backscattering vanishes in
the massless limit: 
\be \label{f(pi)=0}
M=0: \quad f(\pi) = 0 \,.
\ee
The absence of backscattering is a result of
the destructive interference between the time-reversed scattering paths.
This happens since the (pseudo)helicity, the eigenvalue of $\hat \p \bsigma$, 
is asymptotically conserved during scattering off a potential that does not couple 
the Dirac points.
In other words, the Dirac ``spin'' always remains in the 
direction of the particle's momentum.\cite{Ando-Berry} 
The time-reversed backscattering paths then acquire phase difference 
$e^{i\pi}=-1$ that corresponds to the Berry's phase 
$-i\oint\! d\tau \, \psi^\dag\partial_\tau \psi
= \half\oint\! d\tau\, \partial_\tau \theta = \pi$
accumulated while encircling the Dirac point,
$\psi$ being the spinors (\ref{u-plane}) with $M=0$.\cite{Ando-Berry}

\subsection{Born approximation}
\label{sec:born}

Let us now consider the potential term in the Hamiltonian (\ref{H0+U})
as a perturbation, and find the scattering amplitude to the lowest order in $U$.
Following the standard recipe,\cite{Landau3} write the wave function in 
the form 
\be \label{psi=0+1}
\psi = \psi^{(0)} + \psi^{(1)} , 
\ee
where the unperturbed part 
$\psi^{(0)} = e^{i\p\r} u_{\epsilon;\p}$ is a spinor plane wave (\ref{u-plane})
in the direction of the incident momentum $\p$, 
and the scattered part obeys the equation
\be \label{def-psi1}
\lb \HH_0 - \epsilon\rb \psi^{(1)} = - U\psi^{(0)} .
\ee 
The solution of this equation 
\be \label{psi1=}
\psi^{(1)} = - \int\! d^2\r'\, G_\epsilon(\r-\r')
\lb -i\bsigma\bpartial_{\r'} + M\sigma_3 + \epsilon\rb 
U(\r')u_{\epsilon;\p} e^{i\p\r'}
\ee
is found by multiplying both sides 
by the operator $\HH_0 + \epsilon$, and evaluating the operator inverse  
$G_\epsilon = \lb \HH_0^2 - (\epsilon+i0\sign\epsilon)^2\rb^{-1}$
in the Fourier space 
[$\sign\epsilon$ selects the retarded or the advanced part, corresponding to 
the particle and hole sectors]:
\bea \non
G_{\epsilon}(\r) 
&=& \int\! {d^2\k\over (2\pi)^2}\, 
{e^{i\k\r}\over k^2 + M^2 - (\epsilon + i0\sign\epsilon)^2} \\
&=&  {1\over 4\pi} \times  i\pi \sign\epsilon  \times H_0^{(1)}(pr) \,,
\label{G-epsilon}
\eea
where $p= +\sqrt{\epsilon^2 - M^2}$, and
\be \label{hankel}
H_m^{(1,2)}(x) = J_m(x) \pm iY_m(x)
\ee
are the Hankel's functions of the first and second kind.
The asymptotic form of the Green's function (\ref{G-epsilon}) follows 
from Eqs.~(\ref{asympt-large-JY}),
\be \label{G-epsilon-asympt}
G_{\pm|\epsilon|}(\r) \simeq \pm{i\over 4}\sqrt{2\over \pi pr} 
\times e^{ipr - i\pi/4} .
\ee
We now substitute the asymptotic form (\ref{G-epsilon-asympt}) into
Eq.~(\ref{psi1=}), applying the standard approximation
$|\r-\r'| \approx r - \hat \r \cdot \r'$ and
$p|\r-\r'|+\p\r' \approx pr - \q\r'$, where $\p'\equiv p\hat r$ is the
momentum scattered in the direction of observation and $\q=\p'-\p$ 
is the momentum transfer. 
Next we integrate by parts to switch the derivative 
\[
e^{-i\p'\r'} \lb -i\bsigma\bpartial_{\r'} +M\sigma_3 + \epsilon\rb e^{i\p\r'}
\to \lb \bsigma\p' + M\sigma_3 + \epsilon\rb e^{-i\q\r'} 
\]
use
$\lb \bsigma\p'+M\sigma_3+\epsilon\rb u_{\pm|\epsilon|;\p}
=\pm p\, b(\theta)u_{\pm|\epsilon|;\p'}$, and compare the resulting
asymptotic form to Eq.~(\ref{def-f-spinor}) in order to obtain
\bea \label{f-born}
f^{\rm Born}(\theta) &=& -{1\over\hbar v}\sqrt{p\over 8\pi}  
\times  U_\q \, b(\theta) \,, \\
b(\theta) &=& \sqrt{|\epsilon+Mv^2|\over|\epsilon-Mv^2|}
+ e^{-i\theta} \sqrt{|\epsilon-Mv^2|\over|\epsilon+Mv^2|} \,. \quad \quad
\label{b}
\eea
Here $\theta=\angle(\p',\, \p)$, $q=2p\sin(\theta/2)$,
$U_\q = \int\!d\r\, e^{-i\q\r} U(\r)$, and the factor (\ref{b}) 
comes from the spinor structure of the eigenstates (\ref{u-plane}).
We restored $\hbar$ and $v$ so that $\p$ are wave vectors,
with $(\hbar v p)^2 = \epsilon^2 - (Mv^2)^2$, to make it explicit that
$f$ has the dimension of [length]$^{1/2}$.

In the massless limit (pristine graphene), the factor $b(\theta)$ reduces
to the familiar expression coming from the Berry phase,\cite{Ando-Berry} yielding
\be \label{f-born-M=0}
f_{M=0}^{\rm Born}(\theta) = -{1\over\hbar v}\sqrt{p\over 8\pi}  
\times U_\q  \lp 1+e^{-i\theta}\rp \,.
\ee
For $M=0$ the backscattering is absent in agreement with the general property
(\ref{f(pi)=0}).

In the nonrelativistic limit $\epsilon \simeq Mv^2 + (\hbar p)^2/2M$,
both the spinor part $u_\p \to (1 \ 0)^T$ and the Berry phase factor
$b(\theta)\to 2Mv/\hbar p$ become trivial, yielding\cite{Landau3}
\be \label{f-born-M}
f_{\rm nr}^{\rm Born}(\theta) = -{M\over\hbar^2\sqrt{2\pi p}} 
\times U_\q  \,.
\ee

\subsection{Analytical properties}
\label{sec:anal}

\begin{figure}[t]
\includegraphics[width=3in]{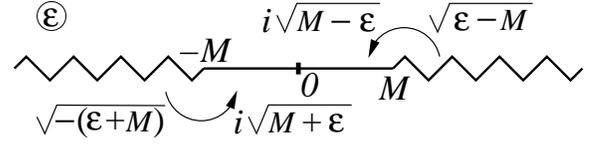}
\caption[]{
Analytic continuation in the energy domain
}
\label{fig:epsilon-cut}
\end{figure}

Here we derive a few properties of the scattering solutions by considering
them as functions of energy $\epsilon$ in the complex plane.\cite{Landau3}
Consider the $r\to \infty$ asymptotic form of the solution to the radial 
equations of Sec.~\ref{sec:rad}, 
\be \label{F-asympt}
F \simeq A(\epsilon) R^+(pr)
+ B(\epsilon) R^-(pr) \,,
\ee
$p(\epsilon)=\sqrt{(\epsilon-M)(\epsilon+M)}$.
Here [cf. Eqs.~(\ref{asympt-large-JY})]
\be \label{JpmiY}
R^\pm(pr) = \sqrt{\pi p\over 2} e^{\pm i\pi j/2} 
H_m^{(1,2)}(pr)
\simeq \frac1{\sqrt{r}} e^{\pm ipr} \,,
\ee
where $H_m^{(1,2)}$ are the Hankel's functions (\ref{hankel}).
The functions $A(\epsilon)$ and $B(\epsilon)$ 
become uniquely defined on the physical sheet of 
the Riemann surface of the square root (Fig.~\ref{fig:epsilon-cut})
described below. On the physical sheet,
the solutions of Eqs.~(\ref{tilde-FG-kappa}) can be obtained from 
those of (\ref{tilde-FG-kappa-dis}) by analytic continuation.

Consider the complex plane of $\epsilon$ (Fig.~\ref{fig:epsilon-cut}) with the 
branch cuts along the real axis connecting the points $\epsilon=\pm M$ with infinity.
The states on the branch cuts correspond to the continuous spectrum, while the real
poles in the interval $-M<\epsilon<M$ correspond to the bound states. 
Define $\sqrt{M-\epsilon}>0$ and $\sqrt{M+\epsilon}>0$ for $-M<\epsilon<M$.
Analytic continuation onto $\epsilon>M$ and $\epsilon<-M$ should agree 
with the standard causality arguments (particles propagate forward in time).
Since the time evolution $\sim \theta(t)e^{-i\epsilon_\p t}$ 
of the particle states ($\Re\epsilon>M$)
is described by the retarded Green's function  
$G^R(\epsilon) \sim (\epsilon-\epsilon_\p+i0)^{-1}$ [here $\theta(t)$ is a unit 
step function], the 
branch cut to the right of $M$ is shifted by the infinitesimal amount $-i0$ below 
the real axis, $\Im\epsilon<0$. The square root $\sqrt{\epsilon-M}$ for $\epsilon>M$
then has to be continued from the upper side of the cut, 
$\sqrt{\epsilon-M} \to i\sqrt{M-\epsilon}$, as in the 
Schr\"odinger case.\cite{Landau3}  Conversely, 
the hole states ($\Re\epsilon<-M$) are governed by the advanced propagator 
$G^A(\epsilon) \sim (\epsilon-\epsilon_\p-i0)^{-1}$, such that 
$\int\!d\epsilon\,G^A(\epsilon) e^{-i\epsilon t}\sim\theta(-t)e^{-i\epsilon_\p t}$, 
effectively shifting the other cut above the real axis, $\Im\epsilon>0$. 
The square root $\sqrt{-(\epsilon+M)}$ for $\epsilon<-M$ then has to be continued
from the lower side of the cut, $\sqrt{-(\epsilon+M)} \to i\sqrt{\epsilon+M}$.
Summarizing,
\bse \label{anal-cont}
\bea
\label{anal-cont+}
\epsilon>0: \quad \sqrt{M-\epsilon} &\to& -i\sqrt{\epsilon-M} \,;\\
\label{anal-cont-}
\epsilon<0: \quad \sqrt{M+\epsilon} &\to& -i\sqrt{-(\epsilon+M)} \,.
\eea
This determines the sign of the continuation 
\be \label{anal-cont-k}
\lambda = \sqrt{M^2-\epsilon^2}\to -ip \,, \quad p = \sqrt{\epsilon^2 - M^2} \,.
\ee
\ese
Note that the $\pm$ signs in front of the square root in Eq.~(\ref{tilde-FG})
[that agree with those in the free spinor (\ref{sph})]
appear naturally as a result of the procedure (\ref{anal-cont}).

The radial functions of the bound states decay for $r\to \infty$. 
This means that the discrete spectrum corresponds to zeros of the
function $B(\epsilon)$ [cf. Eq.~(\ref{JpmiY})].  
The functions $A(\epsilon)$ and $B(\epsilon)$ are connected to 
the partial scattering amplitudes (\ref{S-j}).
Indeed, comparing the asymptotic form (\ref{FG-asymptF}) with (\ref{F-asympt}),
obtain
\be \label{A/B}
A(\epsilon)/B(\epsilon)=e^{2i\delta_j(\epsilon) - i\pi j} \,.
\ee
Thus the amplitude $S_j$ has a pole for any bound state $\epsilon_{\rm b}$.
Following Refs.~\onlinecite{Landau3} and \onlinecite{Landau4}, 
we now express the residue of $S_j$ in this pole 
via the value $A(\epsilon_{\rm b})$.

Consider the radial equations (\ref{FG-kappa}).
Differentiating them with respect to $\epsilon$, obtain
\bea 
\non
(\partial_\epsilon F\sqrt{r})'_r - {\j\over r} \partial_\epsilon F\sqrt{r} 
+ (\epsilon+M-U)\partial_\epsilon G\sqrt{r} &=& -G\sqrt{r}, \quad 
\\
(\partial_\epsilon G\sqrt{r})'_r + {\j\over r} \partial_\epsilon G\sqrt{r} 
- (\epsilon-M-U)\partial_\epsilon F\sqrt{r} &=& F\sqrt{r} . \quad 
\non
\eea
Multiply the first one by $-2\sqrt{r}G$, the second one by $2\sqrt{r}F$,
then multiply Eq.~(\ref{F-kappa}) by $\sqrt{r}G$ and Eq.~(\ref{G-kappa})
by $-\sqrt{r}F$, and add up all the four equations. 
After many terms cancel, what is left can be cast in the following form:
\be \label{r(F^2+G^2)}
\partial_r \lb r \lp F\partial_\epsilon G - G\partial_\epsilon F\rp \rb 
= r(F^2 + G^2) \,.
\ee
Next, integrate (\ref{r(F^2+G^2)}) with respect to $r$ from $r=0$ to $r$,
having in mind the limit $r\to \infty$. The right-hand side becomes unity
due to normalization, while in the left-hand side we use the asymptotic relation
\be \label{F'=G}
(F\sqrt{r})'_r \simeq -(\epsilon+M) G\sqrt{r}
\ee
that follows from Eq.~(\ref{F-kappa}) if one neglects terms with $U(r)$ and $j/r$.
The relation (\ref{F'=G}) allows us to rewrite Eq.~(\ref{r(F^2+G^2)}) 
for $r\to \infty$ in terms of the component $F$ only,
\be \label{F-asympt-rel}
(F\sqrt{r})'_r (\partial_\epsilon F\sqrt{r}) 
- (F\sqrt{r}) (\partial_\epsilon F\sqrt{r})'_r  \simeq \epsilon + M \,.
\ee 
Now consider the asymptotic form (\ref{F-asympt}), where
we set $A(\epsilon)\simeq A(\epsilon_{\rm b})$ and  
$B(\epsilon) \simeq \beta(\epsilon-\epsilon_{\rm b})$, 
$\beta = [\partial B/\partial\epsilon]_{\epsilon=\epsilon_{\rm b}}$.
Substituting it into Eq.~(\ref{F-asympt-rel}), find
\be \label{beta}
\beta = - {1\over 2A(\epsilon_{\rm b})} 
\sqrt{M+\epsilon_{\rm b}\over M-\epsilon_{\rm b}} \,.
\ee
Using (\ref{A/B}), finally obtain the $S$-matrix residue
\be \label{Sj-res}
e^{2i\delta_j(\epsilon)} \simeq -e^{i\pi j} \times
{2[A(\epsilon_{\rm b})]^2\over \epsilon-\epsilon_{\rm b}}
\sqrt{M-\epsilon_{\rm b}\over M+\epsilon_{\rm b}} 
\ee
in terms of the coefficient $A(\epsilon_{\rm b})$ in the asymptotic form
(\ref{F-asympt}) of the wave function.
We will use the result (\ref{Sj-res}) in Sec.~\ref{sec:coul}  
to normalize the bound state wave functions.

\section{Low Energy Scattering}
\label{sec:slow}

As an application of the developed formalism, 
consider scattering off a potential $U(r)$ localized
within the domain of the size $\sim \ell$ much greater than the graphene
lattice constant. 
For concreteness, take $U(r)=V_0 \theta(\ell - r)$.
We will be primarily interested in the situation when 
the range $\ell$ is small compared to the wavelength, $p\ell < 1$. 

First note that at large distances, $r>\ell$, 
where the potential $U(r)$ does not contribute,
the radial components of the spinor (\ref{def-FG}) are linear combinations
of the corresponding free solutions (\ref{sph}) 
\bse \label{FG-JY}
\be \label{F-JY}
F|_{r>\ell} =  C\sqrt{|\epsilon+M|}
\lf J_m(pr) \cos\delta_j  - Y_m(pr) \sin\delta_j\rf , \quad\quad
\ee
\be \label{G-JY}
G|_{r>\ell} = \pm C\sqrt{|\epsilon-M|}
\lf J_{m+1}(pr) \cos\delta_j  - Y_{m+1}(pr) \sin\delta_j\rf ,
\ee
\ese
where $j=m+\half$ and $C=\sqrt{\pi p/|\epsilon|}$.

For $r<\ell$, the regular at $r=0$ solutions take the form
\bse \label{FG-JY-in}
\bea \label{F-JY-in}
F|_{r<\ell} &=&  \widetilde C \sqrt{|\widetilde\epsilon+M|} J_m(\widetilde pr)\,, \\
G|_{r<\ell} &=& \sign\widetilde\epsilon \times
\widetilde C\sqrt{|\widetilde \epsilon-M|} J_{m+1}(\widetilde pr) \,.
\label{G-JY-in}
\eea
Here we introduced $\widetilde \epsilon = \epsilon-V_0$ and 
$\widetilde p \equiv \sqrt{\widetilde \epsilon^2 - M^2}$. 
The solutions (\ref{FG-JY-in}) are written for $|\widetilde\epsilon|>M$.
For $|\widetilde\epsilon|<M$, $\widetilde p \to i\widetilde \lambda$
[cf. Sec.~\ref{sec:anal}, $\epsilon\to \widetilde\epsilon$],
their counterparts read
\bea \label{F-JY-in<}
F|_{r<\ell} &=&  \widetilde C' \sqrt{M+\widetilde\epsilon}\,I_m(\widetilde \lambda r)\,, \\
G|_{r<\ell} &=& -\widetilde C' \sqrt{M-\widetilde\epsilon}\,I_{m+1}(\widetilde \lambda r)\,,
\label{G-JY-in<}
\eea
\ese
where $I_m(x)=i^{-m}J_m(ix)$ is the modified Bessel function of the first kind;
$\widetilde C$ and $\widetilde C'$ are constants that can be chosen real.

Both spinor components must be continuous at $r=\ell$. This translates into 
the following matching condition
\be \label{matching}
\left.{F\over G}\right|_{r<\ell} = \left.{F\over G}\right|_{r>\ell} 
\ee
yielding (for $|\widetilde\epsilon|>M$)
\be \label{matching-detail}
\widetilde\zeta \times {J_m(\widetilde p \ell)\over J_{m+1}(\widetilde p\ell)}
=\zeta \times {J_m(p \ell) -  Y_m(p \ell) \tan\delta_j \over 
J_{m+1}(p\ell)-Y_{m+1}(p \ell) \tan\delta_j} \,,
\ee
where 
\be
\zeta=\sign\epsilon\times \sqrt{\left| {\epsilon+M\over \epsilon-M}\right|}, \quad 
\widetilde\zeta=\sign\widetilde\epsilon\times 
\sqrt{\left| {\widetilde\epsilon+M\over \widetilde\epsilon-M}\right|} .
\ee
For $|\widetilde\epsilon|<M$, 
$\widetilde\zeta \to -i \sqrt{M+\widetilde\epsilon\over M- \widetilde\epsilon}$
[cf. Sec.~\ref{sec:anal}],
and the left-hand side of Eq.~(\ref{matching-detail}) is analytically 
continued to 
$-\sqrt{M+\widetilde\epsilon\over M- \widetilde\epsilon} \times 
I_m(\widetilde\lambda \ell)/I_{m+1}(\widetilde\lambda \ell)$, in accord 
with what one gets by
applying the condition (\ref{matching}) directly to Eqs.~(\ref{F-JY-in<}) and 
(\ref{G-JY-in<}). Hence we will work with Eq.~(\ref{matching-detail})
keeping in mind this analytic continuation for $|\widetilde\epsilon|<M$.

As a result, from Eq.~(\ref{matching-detail}) find
[cf. Eq.~(\ref{hankel})]
\be \label{Sj-slow}
S_j-1=2\times\frac
{\zeta J_{m+1}(\widetilde p\ell)J_m(p\ell)
-\widetilde\zeta J_m(\widetilde p\ell)J_{m+1}(p\ell)}
{\widetilde\zeta J_{m}(\widetilde p\ell)H^{(1)}_{m+1}(p\ell)
-\zeta J_{m+1}(\widetilde p\ell)H^{(1)}_{m}(p\ell)} .
\ee
For the massless case [$\zeta, \widetilde\zeta \to \pm$], Eq.~(\ref{Sj-slow}) 
has been first obtained in Ref.~\onlinecite{katsnelson-novoselov}
(see also Ref.~\onlinecite{hentschel}).

For short-ranged scatterers, $p\ell \ll 1$, the $j=\pm \half$ 
channels provide the main contribution. Indeed, for $m\neq 0,-1$ 
the corresponding $S_j-1$ are small as powers of $p\ell$; this can be seen from 
the short-distance behavior (\ref{bessel-short}).
The scattering amplitude can be thus approximated by 
taking into account only the $j=\pm \half$ channels,
\be \label{f-slow}
f(\theta)\simeq\sqrt{\pi\over 2p}\lb \frac1{{\widetilde\zeta\over\zeta}
{J_0(\widetilde p\ell)\over 
p\ell J_1(\widetilde p\ell)} - \ln {2i\over \gamma_E p\ell}}
+
\frac{e^{-i\theta}}{{\zeta\over\widetilde\zeta}
{J_0(\widetilde p\ell)\over 
p\ell J_1(\widetilde p\ell)} - \ln {2i\over \gamma_E p\ell}}
\rb .
\ee
Here we utilized the asymptotic behavior of the Hankel's functions (\ref{hankel})
\be \label{hankel-01-asympt}
H_0^{(1)}(x) \simeq -{2i\over \pi}\ln{2i\over\gamma_E x} \,, \quad
H_1^{(1)}(x) \simeq {2\over i\pi x} \,.
\ee

In the massless limit, relevant for scattering off short-ranged impurities
in pristine graphene, the $H_1^{(1)}(p\ell)$ contribution dominates,  
and the amplitude (\ref{f-slow}) is small as 
$f\sim p^{-1/2}\times p\ell(1+e^{-i\theta})$, 
\cite{katsnelson-novoselov} resulting in negligible scattering 
away from resonance [$J_0(\widetilde p \ell)\neq 0$].
The angular distribution of the scattered particles has a distinctive
$\cos^2(\theta/2)$ dependence that comes from the spinor structure of 
the eigenstates (\ref{u-plane}), 
and agrees with the general property (\ref{f(pi)=0}).

In the opposite, nonrelativistic limit, $\zeta \gg \widetilde\zeta$,  
the first term ($m=0$ channel) determines the amplitude (\ref{f-slow}),
where now only the logarithmic term coming from $H_0^{(1)}(p\ell)$ 
can be kept in the denominator.\cite{Landau3}


\section{Coulomb scattering}
\label{sec:coul}

\subsection{Short distance behavior. Critical field strength}

The analysis of Sec.~\ref{sec:limit-dirac} 
shows that the low energy Dirac theory is inconsistent 
with singular potentials $U\sim r^{-s}$, $s>1$.  Hence, it is clear
that the Coulomb potential
\be \label{U-coul}
U(r) = - \hbar v \times {\alpha \over r} 
\ee
is a borderline case which should be studied with care: Any slightly more singular
potential at $r\to 0$ would cause the Dirac vacuum breakdown.
Below we consider the potential (\ref{U-coul}) where 
the strength $\alpha$ can be both positive (attraction) 
and negative (repulsion).

Suppose for now that the effective impurity strength
$\alpha$ is sufficiently small, and consider Eqs.~(\ref{FG-kappa}).
Taking the short-distance behavior of the radial wave function 
as $F\sqrt{r} \sim r^\gamma$ and 
$G\sqrt{r} \sim r^\gamma$, and neglecting the non-singular terms as $r\to 0$,
obtain
\be \label{gamma}
\gamma = \sqrt{\j^2 - \alpha^2} \,, \quad 
\j = \pm \ts{\frac12}, \pm \frac32, ... \,.
\ee
The theory (\ref{H0+U}) and (\ref{U-coul}) 
is then consistent when $|\alpha| < |\j|$ for any possible $\j$, 
i.e. under the condition
\be \label{alpha<0.5}
|\alpha| < \ts{\frac12} \,.
\ee
For imaginary $\gamma$, i.e. when $|\alpha|$ exceeds $|\j|$, 
the eigenstates $F$ and $G$ oscillate and have no well-defined limit as $r\to 0$,
which corresponds to the Dirac vacuum breakdown in the same sense as in the 
discussion of Sec.~\ref{sec:limit-dirac}.
Such an upper bound on the potential strength is similar to that in
the nonrelativistic collapse in the $1/r^2$ potential.\cite{Landau3}

The condition (\ref{alpha<0.5}) appears to be even more restrictive than that 
in 3D, where the Dirac theory with a point-like Coulomb potential source 
is consistent for $|\alpha| < 1$.\cite{Landau4}
The problem of what happens when $\alpha>1$ in 3D has been the subject
of intense theoretical investigation.\cite{Pomeranchuk,Gershtein,Popov,ZP}
Classically, this instability corresponds to falling of a K-shell electron
into the potential center.
On a quantum level, the Dirac vacuum breaks down by a sufficiently strong 
Coulomb center with $Z$ above a certain value $Z_c$, 
by creating electron-positron pairs; an electron then binds
to the nucleus while a positron flies off to infinity.
The major difficulty is that the $Z>Z_c$ problem requires ultraviolet 
regularization, such as introducing the finite size of the 
nucleus.\cite{Pomeranchuk} 
However, due to very small value of the fine structure constant
$e^2/\hbar c =1/137$, 
the consequences of this restriction 
never materialized in QED for the K-shell electrons in heavy atoms, as 
$Z\times e^2/\hbar c<1$
for all the known elements in the periodic table, $Z \lesssim 110$.

In a physically relevant case when the field (\ref{U-coul})
is due to a Coulomb impurity in the vicinity of the graphene sheet,
the bare potential strength 
\be \label{alpha}
\alpha_0 = {Z e_*^2\over \hbar v} \,, \quad e_*^2 ={2e^2\over \varepsilon+1} \,.
\ee
Here $Z$ is the impurity valence, $e$ is the unit charge, 
and $\varepsilon$ is the dielectric constant of a substrate.
The vacuum value $\alpha_0|_{Z=1, \varepsilon=1}\approx 2.2$ 
for $v= 1\times 10^6\,$m/s, while for the SiO$_2$ substrate,
$\alpha_0|_{Z=1, \varepsilon=3.9}\approx 0.9$.

Electron-electron interactions result in screening which generally changes the 
shape of the potential. This is what usually happens in a semiconductor 
with a parabolic band, where the Coulomb potential is cut off on the  
screening length scale.
In graphene, due to the semimetallic electron dispersion, the screening is
unusual. In particular, for massless Dirac fermions at half-filling, 
the linear (RPA) screening is scale invariant:\cite{screening,ando'06} 
it preserves the shape of the potential, and simply
reduces the impurity strength, 
\be \label{alpha/RPA}
\alpha_0 \to \alpha = \alpha_0/\varepsilon_{\rm RPA}
\ee
by the factor
\be \label{RPA}
\varepsilon_{\rm RPA} = 1+{q\over 4\hbar v}\times {2\pi e_*^2\over q}
= 1+ {\pi\over 2} \times {e_*^2\over \hbar v} \,.
\ee
Taking literally, the linear screening yields the reduction by the factor
$\varepsilon_{\rm RPA}|_{\varepsilon=3.9}\approx 2.4$
for an impurity strength in the presence of the SiO$_2$ substrate, 
$\alpha|_{Z=1,\varepsilon=3.9}\approx 0.4$.

As one can readily see, due to the threshold (\ref{alpha<0.5}), and a sufficiently
small Fermi velocity $v\approx c/300$,
the situation in graphene is more complex than that in QED. 
For sufficiently large values of $\alpha$, especially 
for multivalent impurities, the nonlinear screening should be applied instead
of the linear (RPA) response, as the latter formally applies only for 
$e_*^2/\hbar v\ll 1$. 
In particular, a practically important question is 
whether the threshold (\ref{alpha<0.5}) 
can at all be determined within the linear screening framework (\ref{RPA}),
i.e. whether it applies to the screened value (\ref{alpha/RPA}).
Strictly speaking, near the threshold, where the bare $\alpha_0\sim 1$, 
the lattice-scale physics starts playing a role, while the RPA dielectric
constant applies in the limit of large distances and weak perturbations. 
The definition of the threshold as  
$|\alpha_0|/\varepsilon_{\rm RPA}<1/2$ may probably be 
used only as an upper estimate of the threshold value.

An even more interesting problem is the screening in  
the massive case.  The linear screening, formally valid for $e^2_*/\hbar v\ll 1$,
is cut off beyond the (reduced) Compton wavelength $\lambda_C = \hbar/Mv$, 
hence the shape of the potential becomes more complex. 
For sufficiently weak interactions, $e_*^2/\hbar v\ll 1$, 
one may argue that the screening can be
neglected, $\alpha \simeq \alpha_0$, at low energies (e.g. for describing 
the bound states), since the corresponding Bohr radius 
$a_B = \lambda_C/(e_*^2/\hbar v) \gg \lambda_C$. 
Taking into account corrections in  
$e_*^2/\hbar v$ would then amount to the ``fine structure'' of the ``atomic levels''
associated with the impurity. On the other hand, for $e_*^2/\hbar v \sim 1$,
the Bohr radius and the Compton wavelength coincide; such a strongly-interacting
``relativistic'' Dirac atom will have deep-lying bound states.
For sufficiently strong potential these states will reach the hole continuum
(critical impurity), resulting in the vacuum breakdown.
In general, this strong-coupling problem, that requires investigation
of the supercritical region, involves many body 
treatment that is beyond the scope of this work.
In what follows we will assume that the condition (\ref{alpha<0.5}) holds
for the effective value of impurity strength $\alpha$, and 
consider only the subcritical regime.

\subsection{Discrete spectrum, $|\epsilon| < M$}

Similar to the 3D case, \cite{Landau4}
we look for the solutions of Eqs.~(\ref{FG-kappa}) in the form [$\hbar =v =1$]
\begin{subequations}
\label{FG-tilde}
\bea
F &=& \sqrt{M+\epsilon} \, e^{-\rho/2} \rho^{\gamma-1/2} \tF(\rho) \,, \\
G &=& \sqrt{M-\epsilon} \, e^{-\rho/2} \rho^{\gamma-1/2} \tG(\rho) \,.
\eea
\end{subequations}
Here $\rho = 2\lambda r$, and $\lambda=\sqrt{M^2-\epsilon^2}$.
After substituting the functions (\ref{FG-tilde}) into (\ref{FG-m}), 
obtain the equations for $\tF$ and $\tG$:
\begin{subequations}
\bea
\rho \tF'_\rho + (\gamma-\j)\tF - {\rho\over 2}(\tF - \tG)
+ {\lambda \alpha \over M+\epsilon} \tG &=& 0 \,, \quad \\
\rho \tG'_\rho + (\gamma+\j)\tG + {\rho\over 2}(\tF - \tG)
- {\lambda \alpha \over M-\epsilon} \tF &=& 0 \,. \quad
\eea
\end{subequations}
Representing 
\be
\tF = Q_1 + Q_2 \quad \mbox{and} \quad \tG = Q_1 - Q_2, 
\ee
find
\bse \label{Q12}
\bea \label{Q12a}
\rho Q_1' + \lp \gamma-{\alpha\epsilon\over\lambda}\rp Q_1 
-\lp \j + {M\alpha\over\lambda}\rp Q_2 &=& 0, \quad\quad\quad \\
\rho Q_2' + \lp \gamma-\rho + {\alpha\epsilon\over\lambda}\rp Q_2 
-\lp \j - {M\alpha\over\lambda}\rp Q_1 &=& 0, \quad\quad\quad
\label{Q12b}
\eea
\ese
from which the equations for $Q_1$ and $Q_2$ are
\bse\label{Q-Kummer}
\bea  \label{Q1}
\rho Q_1'' + (1+2\gamma - \rho)Q_1' 
- \lp \gamma - {\alpha\epsilon\over \lambda}\rp Q_1 &=& 0 \,,  \quad\quad\quad \\
\rho Q_2'' + (1+2\gamma - \rho)Q_2' 
- \lp 1 + \gamma - {\alpha\epsilon\over \lambda}\rp Q_2 &=& 0 \,. \quad\quad\quad
\label{Q2}
\eea
\ese
To derive (\ref{Q-Kummer}) we used the identity 
\be \label{kappa-gamma-id}
\j^2 - M^2\alpha^2/\lambda^2 = \gamma^2 - \alpha^2\epsilon^2/\lambda^2 \,.
\ee
Eqs.~(\ref{Q-Kummer}) are of the Kummer form,
\be \label{hypergeom-eqn}
z\F'' + (c-z)\F' - a\F = 0\,,
\ee
where $\F$ is the confluent hypergeometric function
\be \label{hypergeom}
\F(a, c; z) = 1+{a\over c} {z\over 1!} + {a(a+1)\over c(c+1)} {z^2\over 2!} + ...\,.
\ee
Thus the solutions of Eqs.~(\ref{Q-Kummer})
\bse
\bea 
Q_1 &=& C_1 \F(\gamma-\alpha\epsilon/\lambda, 1+2\gamma; \rho)  \,, \\
Q_2 &=& C_2 \F(1+\gamma-\alpha\epsilon/\lambda, 1+2\gamma; \rho)  \,. 
\eea
\ese
Using $\F(a,c; 0)=1$ and Eqs.~(\ref{Q12}) we find the ratio
\be \label{C12}
c_{12} \equiv {C_2\over C_1} = 
{\gamma - \alpha\epsilon/\lambda \over \j + M\alpha/\lambda} \,,
\ee
and the wave functions of the bound states
\bse \label{FG-bs}
\bea
\non
F = \sqrt{M+\epsilon} \, e^{-\rho/2} \rho^{\gamma-1/2} C_1
\lf \F(\gamma-\alpha\epsilon/\lambda, 1+2\gamma; \rho) 
\right. \\ \left.
+ c_{12}\F(1+\gamma-\alpha\epsilon/\lambda, 1+2\gamma; \rho) \rf , \quad\quad
\label{F-bs}
\\
\non
G = \sqrt{M-\epsilon} \, e^{-\rho/2} \rho^{\gamma-1/2} C_1
\lf \F(\gamma-\alpha\epsilon/\lambda, 1+2\gamma; \rho) 
\right. \\ \left.
- c_{12}\F(1+\gamma-\alpha\epsilon/\lambda, 1+2\gamma; \rho) \rf , \quad\quad
\label{G-bs}
\eea
\ese
where $C_1$ is the overall normalization factor.

Bound states occur when the functions $\F$ reduce to polynomials, 
i.e. when
\be \label{bs-condition}
\gamma(j)-{\alpha\epsilon_{n,j}\over\lambda(\epsilon_{n,j})} = -n \,, \quad 
\begin{cases}
n = 0, 1, 2, ... \ \ \mbox{for $j>0$,}
\\ 
n = 1, 2, 3, ... \ \ \mbox{for  $j<0$.}
\end{cases}
\ee
From Eq.~(\ref{bs-condition}) the bound state energies follow:\cite{khalilov}
\be \label{bs-energy}
\epsilon_{n,j} = {M \sign \alpha \over \sqrt{1+{\alpha^2\over (n+\gamma)^2}}} \,,
\quad \gamma(j)=\sqrt{j^2-\alpha^2} \,.
\ee
The bound states are doubly degenerate, $\epsilon_{n,j}=\epsilon_{n,-j}$.

The overall normalization factor $C_1$ can be found by comparing the $r\to\infty$
asymptotic behavior of the function (\ref{F-bs})
[where the leading contribution comes only from the first term],
with Eq.~(\ref{F-bs-asympt}) that will be obtained below.
The asymptotic behavior of (\ref{F-bs}),
\be \non
F\simeq (-)^n C_1  {\Gamma(1+2\gamma)\sqrt{M+\epsilon}\over \Gamma(1+2\gamma+n)}
(2\lambda r)^{n+\gamma-1/2} e^{-\lambda r}
\ee
is found using the formula
\begin{widetext}
\bea
\F(a,c;z) &=& {\Gamma(c)\over\Gamma(c-a)} (-z)^{-a} {\cal G}(a,a-c+1,-z) 
 + {\Gamma(c)\over\Gamma(a)} e^z z^{a-c} {\cal G}(c-a,1-a,z) \,,
\label{d14}
\\
\label{G-series}
{\cal G}(a,c;z) &=& 1+ {ac\over 1! \times z}+{a(a+1)c(c+1)\over 2! \times z^2}+ ...
\eea
[Ref.~\onlinecite{Landau3}, Eq.~(d.14)].
As a result, the wave functions for the bound states 
[the upper sign corresponds to $F$ and the lower one to $G$; 
$\epsilon\equiv \epsilon_{n,j}$]
\be \label{FG-bs-norm}
\lf \begin{matrix}F \\ G \end{matrix}\rf 
= {(-)^n \lambda^{3/2}\over M \Gamma(1+2\gamma)}
\sqrt{\frac{\Gamma(1+2\gamma+n)(M\pm \epsilon)}{(j+M\alpha/\lambda) \alpha n!}} \, 
e^{-\lambda r} (2\lambda r)^{\gamma-1/2} 
\lf \lp j+M\alpha/\lambda \rp \F(-n, 1+2\gamma; 2\lambda r) 
\mp n \F(1-n, 1+2\gamma; 2\lambda r) \rf .
\ee
The functions (\ref{FG-bs-norm}) are normalized to 
$\int_0^\infty \! r dr\lp F^2 + G^2\rp =1$.

The size $l_{n,j}$ of the bound state wave functions (\ref{FG-bs-norm})
is controlled by the parameter $\lambda=\lambda(\epsilon_{n,j})$, as
\be \label{l-nj}
l(\epsilon_{n,j}) \equiv {1\over \lambda(\epsilon_{n,j})} = 
{\sqrt{(n+\gamma)^2+\alpha^2}\over (Mv/\hbar)\times |\alpha|} 
\equiv \sqrt{(n+\gamma)^2+\alpha^2}\times {a_B\over Z} \,, 
\quad a_B = {\hbar^2 \over M e_*^2} 
\ee
scaling with the ``Bohr radius'' $a_B$ that is a 
ratio of the reduced Compton wavelength $\hbar/Mv$ and the effective 
fine structure constant $e_*^2/\hbar v$ [note again that we assume 
weak coupling $e_*^2/\hbar v\ll 1$].

\subsection{Continuous spectrum, $|\epsilon|>M$}

The simplest way to obtain the continuous spectrum solutions in the problem 
(\ref{H0+U}) and (\ref{U-coul}) is to  
analytically continue the solutions (\ref{F-bs}) and (\ref{G-bs})
according to the procedure (\ref{anal-cont}).
This yields $\rho \to -2ipr$, and the ratio (\ref{C12}) becomes
\be \label{xi}
c_{12} \to e^{-2i\xi_j} = {\gamma - i\alpha_\epsilon \over \j + iM\alpha/p} \,,
\quad \alpha_\epsilon \equiv {\alpha\epsilon\over p} \,.
\ee
The phase $\xi_j$ is real due to the identity (\ref{kappa-gamma-id}).

Consider now the $|\epsilon|<M$ solutions (\ref{FG-bs}). 
The analytic continuation $|\epsilon|<M \to |\epsilon|>M$
to the continuous spectrum, using 
Eqs.~(\ref{anal-cont})  and (\ref{xi}), yields
\bse \label{FG-cont-unnorm}
\bea
F &=& \sqrt{|\epsilon+M|} \, e^{ipr} r^{\gamma-1/2} C_1'
\lf e^{i\xi} \F(\gamma-i\alpha_\epsilon, 1+2\gamma; -2ipr) 
+ e^{-i\xi} \F(1+\gamma-i\alpha_\epsilon, 1+2\gamma; -2ipr) \rf , \quad\quad
\label{F-cont-unnorm}
\\
G &=& \mp i\sqrt{|\epsilon-M|} \, e^{ipr} r^{\gamma-1/2} C_1'
\lf e^{i\xi}\F(\gamma-i\alpha_\epsilon, 1+2\gamma; -2ipr) 
- e^{-i\xi}\F(1+\gamma-i\alpha_\epsilon, 1+2\gamma; -2ipr) \rf , \quad\quad
\label{G-cont-unnorm}
\eea
\ese
where $C_1'$ is some new overall normalization factor that has to be found 
by matching the asymptotic behavior of the solutions (\ref{FG-cont-unnorm})
with Eq.~(\ref{FG-asympt}).
For that, we first consider the asymptotic behavior of the second terms
in Eqs.~(\ref{FG-cont-unnorm}). Using the identity 
[see e.g. Ref.~\onlinecite{Landau3}, Eq.~(d.10)]
\be \label{d10}
\F(a,c;z) = e^z \F(c-a,c,-z) \,,
\ee
we transform 
\be
\F(1+\gamma-i\alpha_\epsilon, 1+2\gamma; -2ipr)  
= 
e^{-2ipr} \lb \F(\gamma-i\alpha_\epsilon, 1+2\gamma; -2ipr) \rb^* .
\ee
As a result, obtain the normalized eigenstates for the continuous spectrum
\begin{subequations}
\label{FG-coul-scatt}
\bea
F &=& {2\over \sqrt{r}} \sqrt{|\epsilon + M| \over 2|\epsilon|} 
 {|\Gamma(1+\gamma+i\alpha_\epsilon)|\over \Gamma(1+2\gamma)} e^{\pi\alpha_\epsilon/2} (2pr)^\gamma
\Re \lf e^{ipr + i\xi} \F(\gamma-i\alpha_\epsilon, 1+2\gamma; -2ipr) \rf ,
\label{F-coul-scatt}
\\
G &=& \pm{2\over \sqrt{r}} \sqrt{|\epsilon- M| \over 2|\epsilon|} 
 {|\Gamma(1+\gamma+i\alpha_\epsilon)|\over \Gamma(1+2\gamma)} e^{\pi\alpha_\epsilon/2} (2pr)^\gamma
\Im \lf e^{ipr + i\xi} \F(\gamma-i\alpha_\epsilon, 1+2\gamma; -2ipr) \rf .
\label{G-coul-scatt}
\eea
\end{subequations}
\end{widetext}
Note that for $\epsilon<-M$, the analytic continuation (\ref{anal-cont-})
produces an extra minus sign for $G$ (here $\pm = \sign \epsilon$), as expected from 
the asymptotic behavior (\ref{FG-asympt}).
One can prove that 
the solutions (\ref{FG-coul-scatt}) are correctly normalized by using the formula
(\ref{d14}). The
asymptotic $pr\to \infty$ behavior of the normalized solutions 
(\ref{FG-coul-scatt})
\be \label{FG-coul-asympt}
\begin{matrix} 
 F  & \simeq &  
{2\over \sqrt{r}}\sqrt{|\epsilon+M|\over 2|\epsilon|}
\cos\lp pr - {j\pi/ 2}  + \alpha_\epsilon \ln 2pr + \delta_j \rp ,
\\
 G  & \simeq & \pm  {2\over \sqrt{r}} \sqrt{|\epsilon-M|\over 2|\epsilon|}
\sin\lp pr - {j\pi/ 2}  + \alpha_\epsilon \ln 2pr + \delta_j \rp 
\end{matrix}
\ee
deviates from that of Eq.~(\ref{FG-asympt}) by the familiar logarithmically
divergent Coulomb phase $\ln (2pr)$, ubiquitous in both 
the nonrelativistic \cite{Landau3,Stern-Howard,Barton} and the
relativistic \cite{Landau4} cases. 
The scattering phases are then
\be \label{delta-coul}
\delta_j = \xi_\j + \frac\pi2 (\j-\gamma) -\arg\Gamma(1+\gamma+i\alpha\epsilon/p) \,,
\ee
with the corresponding $S$-matrix elements (\ref{f})
\be \label{S-coul}
S_j=e^{2i\delta_j} = {\j+iM\alpha/p\over \gamma-i\alpha\epsilon/p} 
{\Gamma(1+\gamma-i\alpha\epsilon/p)\over \Gamma(1+\gamma+i\alpha\epsilon/p)} 
e^{i\pi(\j-\gamma)} \,.
\ee
As expected, the poles of $S_j$ determined by the Gamma-function in the 
numerator of Eq.~(\ref{S-coul}) for $1+\gamma-i\alpha\epsilon/p=1-n$, $n=1,2,...$,
as well as by $\gamma - i\alpha\epsilon/p=0$ for $\j>0$,
that occur for the imaginary $p=i\lambda$,
give the corresponding bound states (\ref{bs-energy}).
The residues at these poles are
\be \label{Sj-res-coul}
S_j \simeq (-)^{n+1}\frac{\lambda^3(j+M\alpha/\lambda)e^{i\pi(j-\gamma)}}
{\alpha M^2 n! \,\Gamma(1+2\gamma+n)(\epsilon-\epsilon_{n,j})} \,.
\ee

We now derive the asymptotic form of the discrete spectrum wave functions
based on the relation (\ref{Sj-res}) between the $S$-matrix residue and
the coefficient $A$ in the asymptotic form of the wave function (\ref{F-asympt}).
In the case of the Coulomb scattering, the coefficient $A$ will itself depend 
on $r$ due to the logarithmically divergent Coulomb phase.
In particular, the left-hand side of Eq.~(\ref{Sj-res}) should be corrected
by the factor 
$e^{2i\alpha_\epsilon \ln 2pr} \to (-)^n e^{i\pi\gamma}(2\lambda r)^{2(n+\gamma)}$.
As a result, near the pole
\be
S_j e^{2i\alpha_\epsilon \ln 2pr} \to -e^{i\pi j} 
\frac{\lambda^3(j+M\alpha/\lambda)(2\lambda r)^{2(n+\gamma)}}
{\alpha M^2 n!\, \Gamma(1+2\gamma+n)(\epsilon-\epsilon_{n,j})} \,,
\ee
which in turn equals the right-hand side of Eq.~(\ref{Sj-res}).
Thus we obtain the asymptotic form 
\begin{align} \label{F-bs-asympt}
F&\simeq {A(r)\over \sqrt{r}}e^{-\lambda r} \,, \quad r\to \infty\,, \\
A(r) &= {\hbar\lambda\over Mv}\sqrt{\frac{(Mv^2+\epsilon)(j+\alpha Mv/\hbar\lambda)}
{2\hbar v\, \alpha \, n!\,\Gamma(1+2\gamma+n)}} 
\times (2\lambda r)^{n+\gamma} . \quad\quad\quad
\non
\end{align}
Here we restored $\hbar$ and $v$, with $\hbar \lambda/v=\sqrt{M^2-\epsilon^2}$, 
so that the dimension of $F$ is explicitly 1/[length]. 
Note also that $|j|<\alpha Mv/\hbar\lambda(\epsilon_{n,j})$, 
so that $A(r)$ is always real.

\subsubsection{Nonrelativistic limit (parabolic band)}
\label{sec:nr-limit}

The nonrelativistic limit occurs when the  ``nonrelativistic velocity''
$v_{\rm nr}$ of the particle is much smaller than the graphene Fermi 
velocity (``speed of light'' $v$) [which we write here explicitly],
\be \label{nr-limit}
v_{\rm nr}\ll v\,, \quad v_{\rm nr} \equiv {\hbar p\over M} \,,\quad
\epsilon_{\rm nr} \equiv \epsilon - Mv^2 \simeq {(\hbar p)^2\over 2M} \ll Mv^2 \,.
\ee
In the limit $v\to\infty$, the fine structure constant
$\alpha\to 0$, whereas the ``nonrelativistic fine structure constant''
\be \label{alpha-nr}
\alpha_{\rm nr} \equiv {Ze_*^2\over \hbar v_{\rm nr}} = {v\over v_{\rm nr}}\alpha 
\ee
remains finite, and determines the Coulomb interaction strength.
In this case
\be
\alpha {\epsilon\over \hbar pv} \to \alpha_{\rm nr} \,, \quad 
\alpha{Mv\over \hbar p} \to \alpha_{\rm nr} \,, 
\quad \gamma \to |\j| = |m|+\half\sign \j \,,
\ee
and using $e^{i\pi(\j-|\j|)} = \sign \j$ and $\Gamma(1+z)=z\Gamma(z)$, 
obtain
\be \label{S-coul-nr}
S_m^{\rm nr} = e^{2i\delta_m^{\rm nr}} =
{\Gamma(|m|+\half - i\alpha_{\rm nr})\over 
\Gamma(|m|+\half + i\alpha_{\rm nr})} \,.
\ee

It is instructive to show by a direct calculation, given below, that 
the phases (\ref{S-coul-nr}) agree with the asymptotic behavior of the 
radial Schr\"odinger wave function $R_m(r)$ of the corresponding nonrelativistic 
Coulomb problem. The radial equation in the presence of the 
Coulomb potential $U(r)\equiv -Ze_*^2/r$ reads
\be \label{rad-schrod}
-\frac1r {d\over dr} \lp r {d\over dr} R\rp + {m^2\over r^2} R
-{2MZe_*^2\over \hbar^2 r} R = {2M \epsilon_{\rm nr}\over \hbar^2} R \,.
\ee
We begin with the bound states. After the substitution
\be \label{Rm-Q}
R = \rho^{|m|} e^{-\rho/2} Q(\rho) \,, \quad \rho = 2\lambda_{\rm nr} r \,,
\quad \hbar\lambda_{\rm nr} = \sqrt{-2M\epsilon_{\rm nr}}
\ee
the problem is reduced to the Kummer equation
\be \label{kummer-nr}
\rho Q'' + (2|m|+1-\rho) Q' - (|m|+\half - \widetilde\alpha)Q = 0 \,,
\ee
where 
\be \label{tilde-alpha-nr}
\widetilde \alpha = \widetilde \alpha(\epsilon_{\rm nr}) 
= {Z\over a_B \lambda_{\rm nr}} \equiv
{MZe_*^2 \over \hbar^2\lambda_{\rm nr}} \,.
\ee
Its solutions
\begin{align} \non 
R &= C_{\rm nr} \times (2\lambda_{\rm nr}r)^{|m|} e^{-\lambda_{\rm nr} r} 
\F(-\widetilde n, 2|m|+1; 2\lambda_{\rm nr} r) , \quad\quad\quad 
\\
-\widetilde n &\equiv |m|+\half -\widetilde\alpha(\epsilon_{\rm nr})\,,
\label{Rm-nr} 
\\
\non
C_{\rm nr} &= (-)^{\widetilde n} 
{\sqrt{2}\lambda_{\rm nr} \sqrt{(2|m|+\widetilde n)!} \over 
(2|m|)!\,\sqrt{\widetilde n! \, (\widetilde n+|m|+\half)}}
\end{align}
become normalizable, $\int_0^\infty\! rdr\, R^2(r)=1$,
with
\be
\F(-\widetilde n, 2|m|+1; 2\lambda_{\rm nr} r) = 
{ (2|m|)! \, \widetilde n! \over [(\widetilde n + 2|m|)!]^2}
\times L_{\widetilde n + 2|m|}^{2|m|}(2\lambda_{\rm nr}r)
\ee
given by the associated Laguerre polynomials,
when $\widetilde n=0,1,2, ...$, yielding
the nonrelativistic spectrum \cite{ZZ}
\be \label{bs-energy-nr}
\epsilon_{\rm nr} = - {Z^2Me_*^4\over 2\hbar^2} \times 
{1\over \lp \widetilde n+|m|+\half \rp^2} \,,
\quad \widetilde n=0,1,2, ...
\ee
in agreement with the corresponding limit of the bound states (\ref{bs-energy}).
[Taking into account the discarded terms $\sim e_*^2/\hbar v$ would 
correspond to the ``fine structure'' of the energy levels (\ref{bs-energy-nr}).]

The normalization coefficient $C_{\rm nr}$ for the bound state solutions 
(\ref{Rm-nr}) is found, as above, via comparing their  
$r\to\infty$ asymptotic behavior with the nonrelativistic 
limit of Eq.~(\ref{F-bs-asympt}),
\be \label{nr-bs-asympt}
R_{r\to\infty} \simeq  
{\sqrt{2}\lambda_{\rm nr} (2\lambda_{\rm nr}r)^{\widetilde n + |m|} \over 
\sqrt{\widetilde n! \, (\widetilde n + 2|m|)! \, (\widetilde n+|m|+\half)}}
\, e^{-\lambda_{\rm nr}r} \,.
\ee
While obtaining Eq.~(\ref{nr-bs-asympt}) we identified
[cf. Eq.~(\ref{bs-condition})]
\be
{\alpha Mv\over \hbar \lambda} \,, {\alpha\epsilon_{n,j}\over \hbar v \lambda} 
\to \widetilde \alpha = n+|j| \equiv  \widetilde n + |m|+\half \,.
\ee

The analytic continuation of the solutions (\ref{Rm-nr}) via 
$\lambda_{\rm nr} \to -ip$, $\widetilde \alpha \to i\alpha_{\rm nr}$, 
and subsequent asymptotic expansion using both terms in 
the formula (\ref{d14}), yields
\be \label{Rm-nr-scatt}
R_m \simeq {2\over\sqrt{r}} 
\cos \lp pr + \alpha_{\rm nr}\ln 2pr +\delta_m^{\rm nr} -\frac{m\pi}2 -\frac\pi4\rp,
\ee
where the scattering phase shifts [defined mod $\pi$]
\be \label{delta-coul-nr}
\delta_m^{\rm nr} = 
- \arg \Gamma(|m|+\half + i\alpha_{\rm nr})
\ee
correspond to the $S$-matrix elements (\ref{S-coul-nr}).

\subsubsection{Ultrarelativistic limit (graphene)}
\label{sec:ultra-lim}

In the ultrarelativistic limit 
$|\epsilon|\gg Mv^2$ 
relevant for pristine graphene monolayer [$M=0$],
\be \label{alpha-eps}
\alpha_\epsilon=\alpha\epsilon/pv \to \alpha \sign\epsilon \,.
\ee
In this case, similar to the 3D Dirac fermions, \cite{Landau4}
the $S$-matrix elements (\ref{S-coul}) become independent of the 
absolute value of energy (depending only on its sign):
\be \label{S-coul-ultra}
e^{2i\delta_j} =  {\j \over \gamma -  i\alpha_\epsilon} 
{\Gamma(1+\gamma - i\alpha_\epsilon)\over 
\Gamma(1+\gamma + i\alpha_\epsilon)} 
e^{i\pi(\j-\gamma)} \,; \quad \delta_j = \delta_{-j} \,.
\ee
Note that the general property (\ref{j=-j}) for massless
fermions holds.

\subsection{Scattering cross section}


We start from the nonrelativistic limit, 
where one can directly sum the series (\ref{f}) with $S_m$ from Eq.~(\ref{S-coul-nr})
to obtain the nonrelativistic scattering amplitude 
in the closed form \cite{Stern-Howard,Barton}
\be \label{f-coul-nr}
f(\theta)=
{-i\over \sqrt{2p \sin^2(\theta/2)}} 
{\Gamma(\half-i\alpha_{\rm nr})\over\Gamma(i\alpha_{\rm nr})}
e^{i\alpha_{\rm nr} \ln \sin^2 (\theta/2)} \,.
\ee
For completeness, the details are given in the Appendix~\ref{app:f-nr}.
Using the property $\Gamma(z)\Gamma(1-z)=\pi/\sin(\pi z)$, the 2D Rutherford
cross section follows\cite{Stern-Howard,Barton}
\be \label{ruth-xsection-nr}
{d\Lambda_{\rm nr} \over d\theta} = 
{\alpha_{\rm nr} \tanh \pi\alpha_{\rm nr} \over 2p \sin^2\frac\theta2} \,.
\ee
Here $\theta$ is the scattering angle, and the momentum transfer 
$q = 2p \sin \frac\theta2$.
The cross section (\ref{ruth-xsection-nr}) is
written in terms of the nonrelativistic fine structure constant 
(\ref{alpha-nr}).


In the opposite, massless limit, 
the phases (\ref{S-coul-ultra}) become independent on the 
absolute magnitude $|\epsilon|$ of energy.
Thus the differential scattering cross section scales with the particle wavelength,
\be \label{sigma-ultra}
\left.{d\Lambda \over d\theta}\right|_{|\epsilon|\gg M} 
= {\tau(\theta) \over |\epsilon|} \,,
\ee
where $\tau(\theta)$ is an $|\epsilon|$-independent function of the scattering angle.
As the symmetry (\ref{j=-j}) is fulfilled, the backscattering 
is absent:
$f(\pi) = 0$ and $\tau(\pi) = 0$, cf. Eq.~(\ref{f(pi)=0}).


For the general case, the differential cross section 
is obtained by summing the series (\ref{f-rel}) with $S_j$ from Eq.~(\ref{S-coul}).
This problem is notoriously cumbersome, as it has long been known 
from the three dimensions.\cite{Landau4}
Unfortunately, for the full relativistic problem $(Mv^2\neq \infty)$, 
neither the differential, 
nor the transport cross section can be obtained in the closed form.
Moreover, the series (\ref{f-rel}) for the total cross section
with the phase shifts (\ref{S-coul}) does not converge. 
To obtain the converging expression for the scattering amplitude, 
one needs to perform an appropriate resummation of this series.
\cite{Landau4,Gluckstern-Lin}
In Appendix~\ref{app:f-gen} we show how to represent the Rutherford scattering
amplitude via the convergent double integral, where the energy and 
angular dependences are separated.
In the following section we numerically sum the series (\ref{lambda-tr-gen})
for the transport cross section.


\section{Exact Coulomb transport cross section in graphene}
\label{sec:transport}

For a half-filled $\pi$-electron band, the RPA screening 
(\ref{alpha/RPA}) is scale-invariant, preserving the {\it functional form} 
of the potential. This is why the exact solution 
for the Coulomb potential (\ref{U-coul}) can be practically important,
as it survives the interaction effects at least on the linear screening level.

\begin{figure}[t]
\begin{minipage}[t]{3.5in}
\includegraphics[width=3.5in]{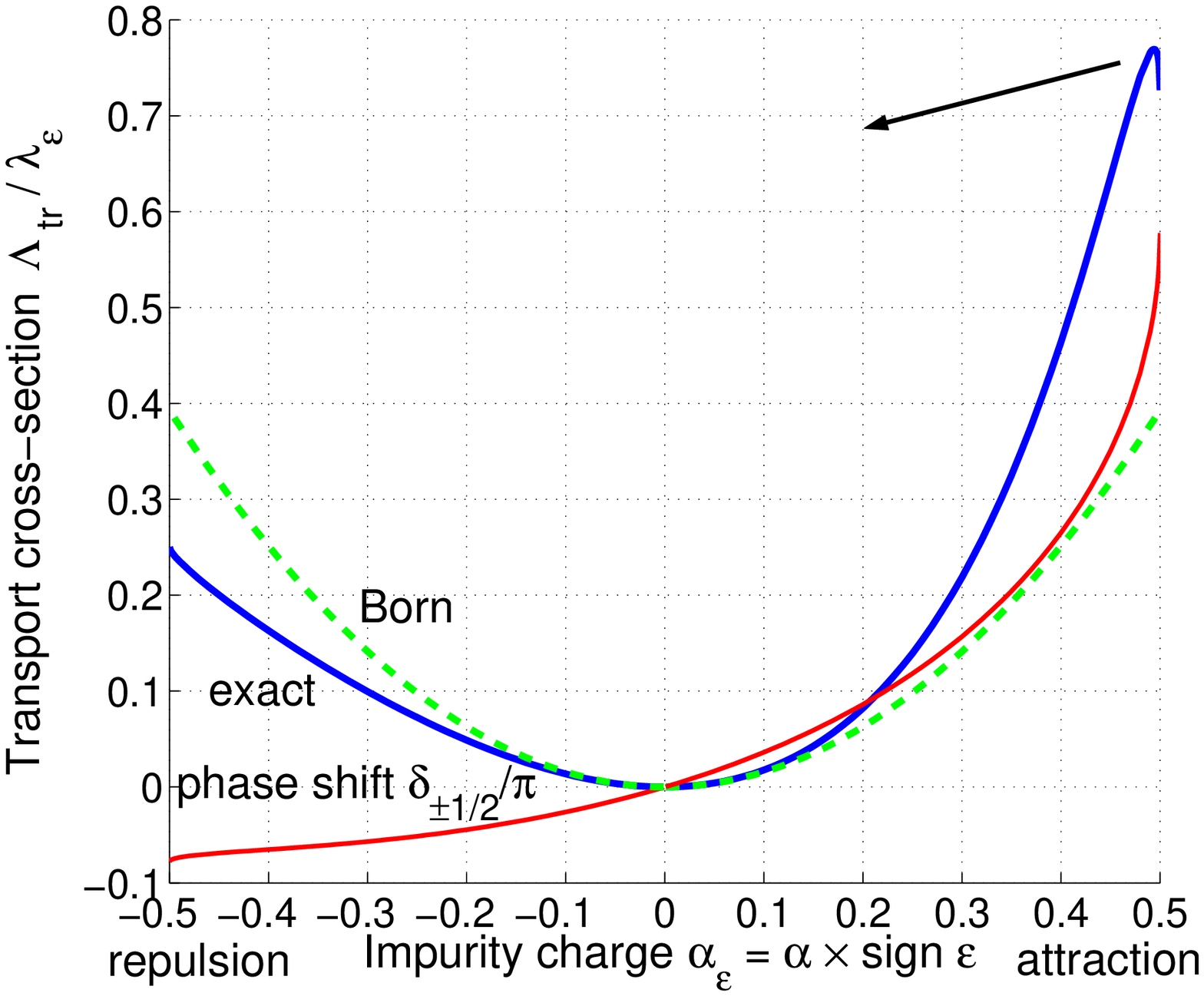}
\end{minipage}
\begin{minipage}[t]{1.6in}
\vspace{-2.8in}\hspace{-0.3in}
\includegraphics[width=1.6in]{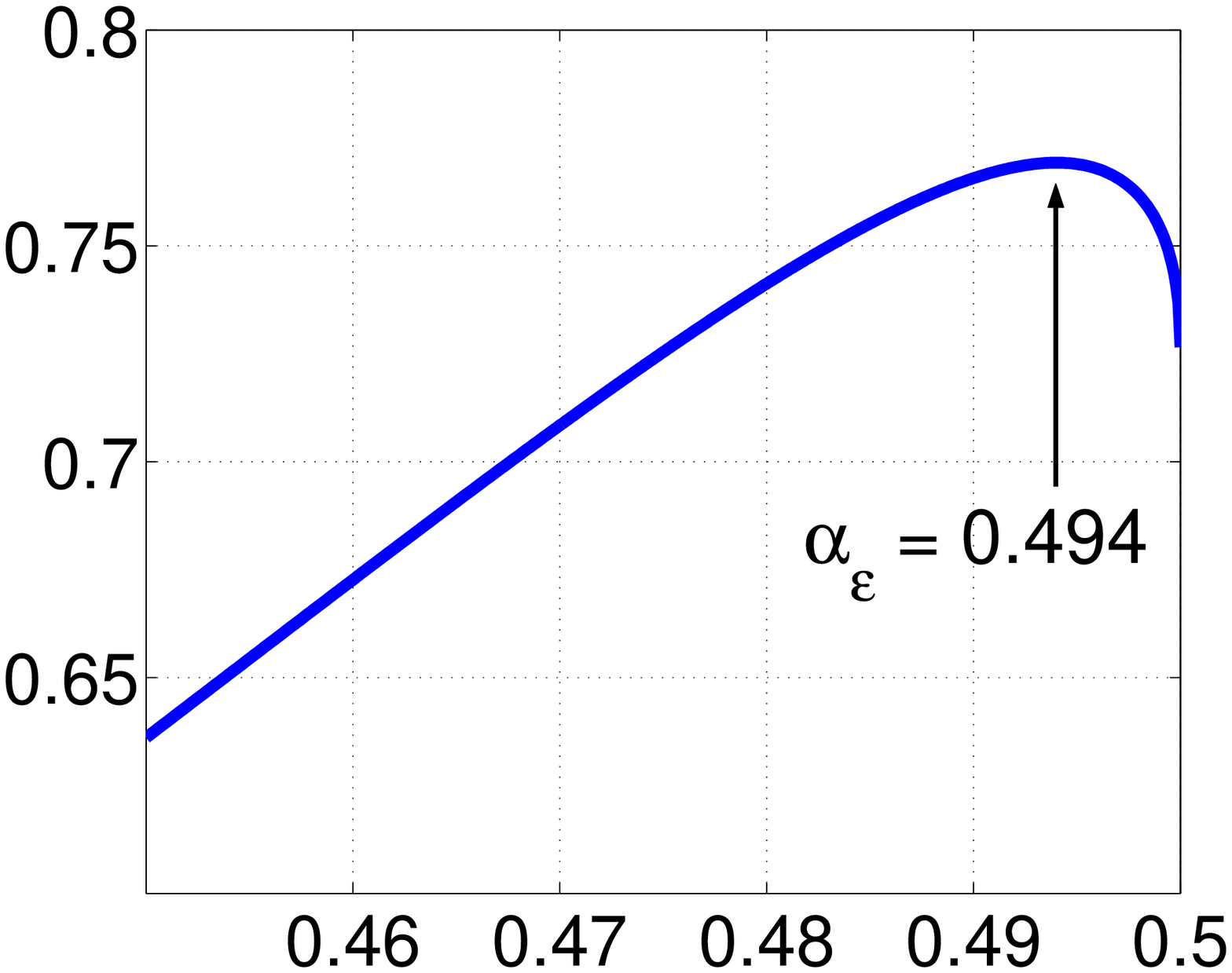}
\end{minipage}
\caption{(Color online) 
Transport cross section as a function of the impurity charge $\alpha$
(solid line). 
Dashed line is the Born approximation; thin red line is the lowest 
scattering phase shift $\delta_{\pm 1/2}/\pi$ for $|j|=1/2$.
Note that $\delta_{\pm 1/2} = \pi/2$ (the unitary limit)
for $\alpha \sign \epsilon = \alpha_\epsilon^* \approx 0.494$; 
the transport cross section {\it decreases} for 
$\alpha_\epsilon^* < \alpha_\epsilon < 1/2$, as shown in the inset.
Attraction here means mutual attraction between the charged carrier and 
the impurity, i.e. $\sign\alpha=\sign\epsilon$, while repulsion stands 
for $\sign\alpha=-\sign\epsilon$.
  }
\label{fig:transp}
\end{figure}

The knowledge of the scattering phases (\ref{S-coul-ultra}) allows us
to obtain the exact transport cross section (\ref{lambda-tr-gen})
for scattering off the Coulomb impurity located in the immediate vicinity of 
the graphene sheet.
Since the phase shifts are energy independent for $M=0$, the transport cross section
is proportional to the energy-dependent carrier wavelength $\lambda_\epsilon$,
\be \label{lambda-tr-C}
\Lambda_{\rm tr} = C(\alpha_\epsilon) \times \lambda_\epsilon \,, 
\quad \lambda_\epsilon = {2\pi\hbar v/|\epsilon|} \,.
\ee
The dimensionless function $C(\alpha_\epsilon)$
[$\alpha_\epsilon=\alpha \sign\epsilon$],
which is the transport cross section in the units of the carrier wavelength, 
is plotted in Fig.~\ref{fig:transp}.

The transport cross section (\ref{lambda-tr-C}) has a few distinct features:
(i) it is {\it not symmetric} with respect to the sign of the potential
as seen by the carrier: A  positively charged impurity ($\alpha>0$)
scatters conduction electrons ($\epsilon>0$) more effectively 
than it scatters holes $(\epsilon<0)$.
The scattering asymmetry with respect to the sign of the potential
arises naturally [e.g. in the next-to-leading Born approximation for low-enegy
particles, Problem 6 in Sec.~132 of Ref.~\onlinecite{Landau3}].
Physically, one may expect the particle to spend more time around an attractive
potential center and thereby be more significantly deflected
(although for an ultrarelativistic particle this intuition may fail).
However, for the practically important 2D and 3D Coulomb scattering in a parabolic
band, the corresponding exact solutions are somewhat exceptional in a sense
that they lack such an asymmetry. Remarkably, for the ``relativistic'' carrier
dispersion, characteristic of graphene, this generally expected asymmetry
is recovered. 
(ii) The cross section is apparently {\it non-monotonic} 
for the attractive Coulomb scatterers.
(iii) The {\it unitary limit}, $\delta_{|j|=1/2}=\pi/2$ 
for the $j=\pm\half$ partial wave, is reached for the mutual attraction when
\be \label{alpha*}
\alpha^*_\epsilon \equiv \alpha \times \sign\epsilon\approx 0.494\,,
\ee
just below criticality.

We were not able to link the above unitarity to any resonance or other
special behavior at the point (\ref{alpha*}), which may as well be accidental.
(One may argue that after subtracting the logarithmically divergent Coulomb
phase, the phase shifts alone have lost their meaning, whereas the differences
between them correspond to observable quantities.) 
Indeed, the partial Coulomb scattering phases $\delta_j$ even in the nonrelativistic 
Rutherford problem \cite{Landau3} generally pass the value $\pi/2$
at non-special values of parameters. At that point, the particular angular momentum 
channel reaches unitarity (maximum possible scattering). 
However, in previously studied cases this unitarity
in one channel did not cause a local maximum for
the sum (\ref{sigma-tr}) over all channels. 
The opposite situation apparently 
happens in the 2D Dirac case: the relatively strong dependence
of the scattering cross section on the lowest-$j$ phase shift causes
the local maximum shown in Fig.~\ref{fig:transp} inset.

The conductivity of graphene monolayer in the presence of charged impurities
with the transport cross section (\ref{lambda-tr-C}) 
is obtained in Ref.~\onlinecite{graphene-asym}. The attraction-repulsion asymmetry
of the cross section can in principle allow one to determine the numbers of 
positively and negatively charged impurities independently.



\subsection*{Born approximation}

We would now like to compare the exact result obtained 
above with the previously used 
Born approximation.\cite{ando'06,nomura,hwang-adam-dassarma}
The Born scattering amplitude (\ref{f-born-M=0}) is straightforwardly found
using $U_\q = -2\pi \hbar v \alpha/q$:
\be \label{f-born-coul}
f^{\rm Born}(\theta) = \alpha \sqrt{\pi\over 8p} \times 
{1+e^{-i\theta}\over \sin(\theta/2)} \,.
\ee
Thus the differential cross section
\be \label{Lambda-born-coul}
{d\Lambda^{\rm Born}\over d\theta} 
= {\pi \alpha^2\over 2p}\times \cot^2{\theta\over 2} \,,
\ee
and the transport cross section
\be \label{xsect-born}
\Lambda_{\rm tr}^{\rm Born} = {\pi^2\alpha^2\over p}
\equiv C^{\rm Born}(\alpha) \times \lambda_\epsilon \,,
\quad C^{\rm Born} = \half \pi\alpha^2 .
\ee
Note that the differential cross section is singular for $\theta=0$ 
as is expected from the long-range character of the Coulomb field.

The above cross sections can also be obtained from the Golden Rule. 
For completeness, we present such a calculation for the 
momentum relaxation time in the presence of $n_i$ Coulomb 
impurities per unit area:
\bea \non
{\hbar \over \tau_{\rm tr}^{\rm Born}(\epsilon)} 
&=& 2\pi n_i \int\! {d^2\p'\over (2\pi\hbar)^2} 
|{\cal M}_{\p\p'}|^2 \delta(\epsilon_\p - \epsilon_{\p'})
(1-\cos\theta) \\
&=& {n_i \pi^2 (Ze_*^2)^2 /\epsilon} \to
{n_i \pi^2 (\hbar v \alpha)^2 /\epsilon} \,,
\label{tau-born}
\eea
where the intra-band matrix element of the interaction
taken between the spinor plane wave states (\ref{u-plane})
\be 
{\cal M}_{\p\p'}
=\int\! d\r\, \psi^\dag_{\epsilon;\p'} U(r) \psi^\pdag_{\epsilon;\p}
= \half \lp 1 + e^{i\theta}\rp U_\q \,.
\ee
Here 
the Fourier transform of the effective potential (\ref{U-coul}) is 
$U_\q = -2\pi Ze_*^2/q$, and in Eq.~(\ref{tau-born}) we accounted for RPA screening
via the procedure (\ref{alpha/RPA}).
The transport cross section (\ref{xsect-born}) follows from 
$n_i v\tau_{\rm tr}\Lambda_{\rm tr} = 1$.
We remark in passing on the following curious observation 
specific for the Born approximation: The ultrarelativistic 
Coulomb transport rate (\ref{tau-born}) written in terms of the quasiparticle 
energy $\epsilon$ formally coincides with the corresponding 
nonrelativistic value in two dimensions, where $\epsilon = p^2/2m^*$.

We also note that, as it is generally expected for the potential 
scattering,\cite{Landau3} the Born approximation (\ref{xsect-born})
overestimates the exact result for the repulsion and underestimates it 
for the attraction. Fig.~\ref{fig:transp} shows that, 
numerically, Born approximation works well for 
$\alpha \lesssim 0.1$, while for the experimentally relevant values
$\alpha\simeq 0.5$ it fails by about a factor of two. 
On a qualitative level, 
since the Born scattering assumes small phase shifts, it fails to recognize 
the strong repulsion/attraction asymmetry,
the unitary scattering occurring at the value (\ref{alpha*}),
and the associated non-monotonic
dependence of the cross section for the mutually attracting carrier and 
impurity.

\section*{\uppercase{Summary}}

In this work we have outlined in detail 
the elastic scattering theory for the (2+1)-dimensional 
massive Dirac fermions in an axially-symmetric potential.
The formalism is relevant for the transport in pristine graphene monolayers 
(massless limit), and for graphene layers with the broken symmetry
between the sublattices (resulting in the finite Dirac mass gap), in the presence of
a smooth potential disorder. 
We showed that the Dirac theory becomes sensitive
to the lattice scale for the potentials that are more singular than $1/r$
as $r\to 0$. We also considered scattering off a localized potential
whose size is smaller than the Dirac fermion wavelength 
(but larger than the graphene lattice scale). 
For the Coulomb scattering, $U=-\hbar v \alpha/r$,
the exact solution is found below the threshold $|\alpha|<1/2$;
from the scattering phase shifts we obtain the exact transport 
cross section.  The transport cross section for the massless case (pristine graphene)
is shown to exhibit a pronounced asymmetry with respect to 
attraction versus repulsion between the charge carrier and the Coulomb impurity.

{\it Note added:} (i) Upon completion of this work we learned about  
the preprint\cite{levitov-shit} where, for the Coulomb potential,
the opposite, supercritical situation $|\alpha|>1/2$
is discussed for the massless limit.
The perturbative renormalization group treatment of Ref.~\onlinecite{levitov-shit},
based on the single-particle Friedel sum rule, 
applies in the weakly-interacting limit $e_*^2/\hbar v\ll 1$.
The strong coupling limit of the problem, valid for $e_*^2/\hbar v\sim 1$
and large impurity charge $Z\gg 1$, was subsequently considered in 
Ref.~\onlinecite{FNS}, yielding a qualitative change in the 
screened potential profile. 
(ii) Also, recently, angular resolved photoemission spectroscopy 
measurement\cite{lanzara-gap} became available, according to which 
the Dirac gap $\Delta \sim 0.1\,$eV opens up for graphene on SiC substrate.
In the presence of Coulomb impurities this would lead to the subgap 
states (\ref{bs-energy}) with the wave function spread over $\sim 10\,$nm
[cf. Eq.~(\ref{l-nj})]. Localized states on this length scale can be detected 
using scanning techniques.

\section*{\uppercase{Acknowledgments}}

It is a pleasure to thank L. Glazman, M. Voloshin,  
B. Shklovskii, and A. Shytov for helpful discussions.
This work was supported by NSF MRSEC grant DMR 02-13706 (at Princeton)
and NSF grants DMR 02-37296 and DMR 04-39026 (at FTPI).

\appendix

\begin{widetext}

\section{Nonrelativistic Coulomb scattering amplitude}
\label{app:f-nr}

Here we show how to sum the series (\ref{f}) with $S_m$ from Eq.~(\ref{S-coul-nr}).
For that, first use the integral representation
\be
\frac{\Gamma(m+\half -i\alpha_{\rm nr})}{\Gamma(m+\half+i\alpha_{\rm nr})}
= {\B(m+\half-i\alpha_{\rm nr}, 2i\alpha_{\rm nr})\over \Gamma(2i\alpha_{\rm nr})}
= {1\over \Gamma(2i\alpha_{\rm nr})} \int_0^1 \! dt\, 
t^{m-1/2-i\alpha_{\rm nr}} (1-t)^{2i\alpha_{\rm nr}-1} \,.
\ee
Noting that $S^{\rm nr}_{-m}=S^{\rm nr}_m$, we then sum the two 
similar looking geometric series. 
Defining $z=e^{i\theta}$, find
\be \label{two-series-nr}
\sum_{m=0}^\infty S_m z^m +
\sum_{m=1}^\infty S_m z^{-m} 
= 
{\Gamma(\half-i\alpha_{\rm nr})\over \Gamma(\half+i\alpha_{\rm nr})}
\widetilde\F(1,\half-i\alpha_{\rm nr},\half+i\alpha_{\rm nr}; z) 
+
\ds{{\Gamma}(\ts{\frac32}-i\alpha_{\rm nr})\over 
{\Gamma}(\ts{\frac32}+i\alpha_{\rm nr})}
{1\over z} 
\widetilde\F(1,\ts{\frac32}-i\alpha_{\rm nr},\frac32+i\alpha_{\rm nr}; z^{-1}) \,. 
\ee
Here we used the integral representation [Ref.~\onlinecite{GR}, Eq.~(9.111)]
\be \label{2F1-int}
\widetilde\F(a,b,c;z) = 
{1\over \B(b,c-b)}
\int_0^1 \! dt\, t^{b-1} (1-t)^{c-b-1} (1-zt)^{-a} 
\ee
for the hypergeometric function 
\be \label{2F1}
\widetilde\F(a,b,c;z) 
= 1 + {ab\over c}{z\over 1!} + {a(a+1) b(b+1)\over c(c+1)}{z^2\over 2!} + ... \,.
\ee
Now use the analytic continuation of the hypergeometric series for $|z|>1$, 
Eq.~(9.132.2) of Ref.~\onlinecite{GR} [Eq.~(e.6) of Ref.~\onlinecite{Landau3}],
\be \label{e6}
\widetilde\F(a,b,c;z)=
{\Gamma(c)\Gamma(b-a)\over \Gamma(b)\Gamma(c-a)}
(-z)^{-a}\widetilde \F(a,a+1-c,a+1-b;z^{-1}) 
+ {\Gamma(c)\Gamma(a-b)\over \Gamma(a)\Gamma(c-b)}
(-z)^{-b}\widetilde \F(b,b+1-c,b+1-a;z^{-1})
\ee
to transform the second term of Eq.~(\ref{two-series-nr}) to become the function
of $z$ rather than $z^{-1}$.
Application of the first term of the formula (\ref{e6}) 
cancels the first term of the sum (\ref{two-series-nr}), while the second term
of Eq.~(\ref{e6}) yields
\be
\sum_{m=-\infty}^\infty S_m z^m = 
-{\Gamma(-\half+i\alpha_{\rm nr})\Gamma(\ts{\frac32}-i\alpha_{\rm nr}) 
\over \Gamma(2i\alpha_{\rm nr})}
(-z^{-1})^{-1/2+i\alpha_{\rm nr}} 
\widetilde\F (\ts{\frac32}-i\alpha_{\rm nr},1-2i\alpha_{\rm nr},
\frac32-i\alpha_{\rm nr};z) \,.
\ee
Finally, using Eq.~(9.131) of Ref.~\onlinecite{GR} 
[Eq.~(e.4) of Ref.~\onlinecite{Landau3}],
\be \label{e4}
\widetilde\F(a,b,c;z) = (1-z)^{c-a-b} \widetilde\F(c-a,c-b,c;z) \,,
\ee 
as well as the doubling formula [Ref.~\onlinecite{GR}, Eq.~(8.335.1)]
\be \label{Gamma(2z)}
\Gamma(2x)={2^{2x-1}\over \sqrt{\pi}} \Gamma(x)\Gamma(x+\half) \,,
\ee
and $-(1-z)^2/z = 4\sin^2(\theta/2)$,
one obtains the scattering amplitude
(\ref{f-coul-nr}).


\section{Relativistic Coulomb scattering amplitude}
\label{app:f-gen}

Consider the following transformations of the series
for the scattering amplitude (\ref{f-rel}), $z\equiv e^{i\theta}$: 
\bse
\label{sum-transform}
\bea \label{sum-transform-1}
\sum_{m=-\infty}^\infty S_m z^m 
&=& 
z^{-1/2} \sum_{\ds{\j=\pm \half, \pm \ts{\frac32}, ...}} 
z^\j (\j + i\widetilde{M}) 
{\Gamma(\gamma-i\alpha_\epsilon)\over \Gamma(1+\gamma+i\alpha_\epsilon)} e^{i\pi(\j-\gamma)}
\\   \label{sum-transform-2}
&=&
{z^{-1/2}\over \Gamma(1+2i\alpha_\epsilon)}
\int_0^1 \! dt\, t^{-i\alpha_\epsilon} (1-t)^{2i\alpha_\epsilon} 
\int_{\cal C}\! {d\kappa\over 2\pi i}\, 
\pi\tan (\pi\kappa)\,
(\kappa+i\widetilde{M}) z^\kappa t^{\gamma-1} 
e^{i\pi(\kappa-\gamma)} 
\\   \label{sum-transform-3}
&=&
-{z^{-1/2}\over \Gamma(1+2i\alpha_\epsilon)}
\int_0^1 \! dt\, {d\over dt} \lf t^{-i\alpha_\epsilon} (1-t)^{2i\alpha_\epsilon}\rf 
\int_{\cal C'}\! {d\kappa\over 2\pi i}\, 
{\pi\tan (\pi\kappa)\over \gamma(\kappa)}\,
(\kappa+i\widetilde{M}) z^\kappa t^{\gamma} 
e^{i\pi(\kappa-\gamma)} \,.
\eea
\ese
\end{widetext}
Here $\widetilde{M} \equiv \alpha M/p$, and in the first line we canceled the 
denominator $\gamma-i\alpha_\epsilon$ by utilizing the property $\Gamma(x+1)=x\Gamma(x)$.

\begin{figure}[t]
\includegraphics[width=3.5in]{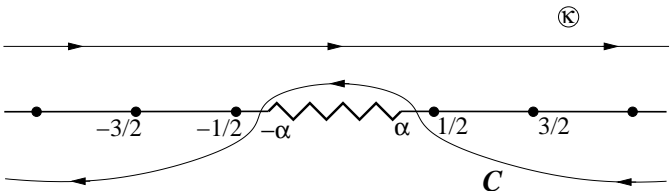}
\caption[]{
The contour ${\cal C}$ of integration in the complex $\kappa$-plane
for summing over $\kappa=\pm 1/2, \pm 3/2, ...$  
}
\label{fig:C}
\end{figure}

\begin{figure}[t]
\includegraphics[width=2.3in]{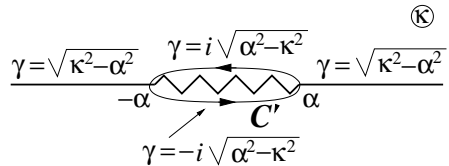}
\caption[]{
The final contour ${\cal C}'$ of integration in the complex $\kappa$-plane 
}
\label{fig:gamma}
\end{figure}

In the next line, Eq.~(\ref{sum-transform-2}), we utilize the integral
representation for the Euler's Beta function 
\be \label{B}
\B(p,q)={\Gamma(p)\Gamma(q)\over \Gamma(p+q)} 
= \int_0^1\! dt\, t^{p-1}(1-t)^{q-1} \,.
\ee
The summation is reduced to the contour integration by virtue of  
the fact that $\pi \tan \pi \kappa$ has residues at the points 
$\kappa=\pm 1/2, \pm 3/2, ...$ with a value $-1$. 
The $\kappa$-integration is along the contour ${\cal C}$ shown in Fig.~\ref{fig:C}.

Finally, in the line (\ref{sum-transform-3}) 
we have performed the integration by parts in the $t$-variable, 
$t^{\gamma-1} = d(t^\gamma)/\gamma$. This way all the remaining singular
behavior in the $\kappa$-plane 
[besides the residues at $\kappa=\pm 1/2, \pm 3/2, ...$]
is reduced to that of $\gamma(\kappa)$ in the denominator. 
The phase of the square root in $\gamma(\kappa)=\sqrt{\kappa^2 - \alpha^2}$
is defined in a standard way, Fig.~\ref{fig:gamma}.
The exponential function $e^{i\pi(\kappa-\gamma)}$ does not diverge for
large imaginary $\kappa$, which allows us to deform the integration
contour ${\cal C}$ in Eq.~(\ref{sum-transform}) 
to ${\cal C}'$ around the cut between 
$\kappa = \pm \alpha$ shown in Fig.~\ref{fig:gamma}. 
Clearly, this procedure is valid only if the condition (\ref{alpha<0.5}) is 
satisfied.



\end{document}